\documentclass[journal]{IEEEtran}

\usepackage[caption=false]{subfig}
\usepackage{amsmath,amsfonts}
\usepackage{array}
\usepackage{textcomp}
\usepackage{stfloats}
\usepackage{url}
\usepackage{verbatim}
\usepackage{graphicx}
\usepackage{amsthm} 
\usepackage{multirow}
\usepackage{graphicx,color}

\usepackage{epstopdf}
\usepackage{epsfig}
\usepackage{setspace}
\usepackage{float}
\usepackage{graphicx}
\usepackage{url}
\usepackage{rotating} 
\usepackage{longtable}
\usepackage{wasysym}
\usepackage[bookmarksopen,bookmarksnumbered,colorlinks,breaklinks,citecolor=red,linkcolor=blue, hyperfootnotes=false]{hyperref} 

\begin{document}
\title{Channel Access Methods for RF-Powered IoT Networks: A Survey}
\author{Hang Yu$^\dag$, Lei Zhang$^\dag$, Yiwei Li$^\dag$, Kwan-Wu Chin and Changlin Yang.
\thanks{Author Yu is with the School of Computer Science, Nanjing University of Information Science and Technology.  Email: hangyu@nuist.edu.cn.
Author Zhang is with the school of Electronic Engineering, North China University of Water Resources and Electric.   Email: zl85370446@sina.com.
Author Li is with the School of Information Engineering, Henan University of Science and Technology.  Email: yl743@outlook.com.
($^\dag$First three authors contributed equally).
Author Chin is with the School of Electrical, Computer and Telecommunications Engineering, University of Wollongong.  Email: kwanwu@uow.edu.au
Author Yang (correspondent author) is with the School of Software Engineering, Sun Yat-Sen University, Zhuhai, China. 
Email: yangchlin6@mail.sysu.edu.cn.}}
\maketitle

%
\begin{abstract}
Many Internet of Things (IoT) networks with Radio Frequency (RF) powered devices operate over a shared medium.  They thus require a channel access protocol.
Unlike conventional networks where devices have unlimited energy, in an RF-powered IoT network, devices must first harvest RF energy in order to transmit or/and receive data. 
To this end, this survey presents the {\em first} comprehensive review of prior works that employ contention-based and contention-free protocols in IoT networks with one or more {\em dedicated} energy sources.  Specifically, these protocols work in conjunction with RF-energy sources to deliver energy delivery or/and data.
In this respect, this survey covers protocols based on Aloha, Carrier Sense Multiple Access (CSMA), polling, and dynamic Time Division Multiple Access (TDMA).  Further, it covers successive interference cancellation protocols.
It highlights key issues and challenges addressed by prior works, and provides a qualitative comparison of these works.
Lastly, it identifies gaps in the literature and presents a list of future research directions.
\end{abstract}

\IEEEpeerreviewmaketitle

\section{Introduction}
\begin{figure}[htbp] 
    \centering 
    \includegraphics[width=0.3\textwidth]{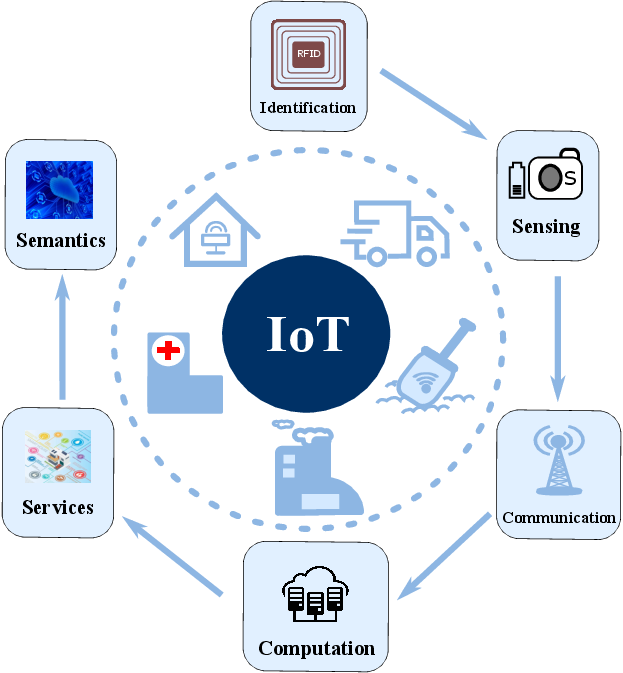} 
    \caption{Six main elements of IoT. In general, sensor devices gather data, which may process data locally or offload it to a gateway.  Processed data is then provided to end users. }
    \label{fig:iotelement} 
\end{figure}

Internet of Things (IoT) networks consist of objects (things) instrumented with a Radio Frequency Identification (RFID) tag, processor, transceiver, sensors, or/and actuators.   These objects/things can be {\em static}, e.g., devices that monitor the ingresses/egresses of buildings~\cite{6676772}, or {\em mobile}, e.g., vehicles such as cars or drones~\cite{8570043}.
An overview of key technologies used in IoT is illustrated in Figure~\ref{fig:iotelement}.  They include identification, sensing, communication, computation, services and semantics; see Table~\ref{table:iottech} for example technologies of each element. 
These technologies in turn lead to the creation of smart cities~\cite{8675165} and factories~\cite{6742605}.  Moreover, IoT is critical to Metaverse, digital twins, and extended reality systems among others~\cite{10105150}.

\begin{table}[!htbp]
\caption{Example IoT technologies for each element shown in Figure~\ref{fig:iotelement}.}\label{table:iottech}
\footnotesize
\centering
\begin{tabular}{|l|l|}
\hline
\textbf{IoT element} & \textbf{Technology}\\
\hline
\hline
{\em Identification} & Electronic product code (EPC), ubiquitous codes \\
                                                & (uCode)~\cite{koshizuka2010ubiquitous}.\\
\hline
{\em Sensing} &  RFID or NFC tags, barcode, QR code, sensor \\
              & networks~\cite{shahraki2021a}.\\
\hline
{\em Communication} & 6LoWPAN~\cite{hui2010ipv6}, WiFi~\cite{wang2016access}, Z-Wave~\cite{gomez2010wireless} ZigBee~\cite{de2016zigbee}, \\
 & NB-IoT~\cite{akpakwu2018a}.\\
\hline
{\em Computation} & Processor, e.g., Arduino, Intel Galieo, Raspberry Pi; \\ 
 &  Operating System (OS), e.g., Android, Contiki,  \\ 
 & TinyOS; cloud platform, e.g., Nimbits, Hadoop~\cite{al2015internet}.\\
\hline
{\em Services} &  Data storage, management and analysis~\cite{rafique2020com}, and \\
              & security~\cite{ alwarafy2021a}, \\               
\hline
{\em Semantic} & Resource Description Framework (RDF), the Web  \\
               & Ontology Language (OWL), Efficient XML \\        
               & Interchange (EXI)~\cite{lerche2012industry}.\\
\hline
\end{tabular}
\end{table}

%
%
Another key technology adopted by IoT systems/networks is Energy Harvesting (EH).   Specifically, devices may harvest energy from environmental sources such as solar~\cite{hamza2021hybrid}, wind~\cite{tan2011optimized}, piezoelectric~\cite{oh2017powering}, kinetic~\cite{magno2018micro} and thermal~\cite{leonov2013ther}, see Table~\ref{table:ehsource} for a summary. 
These technologies in turn help prolong the lifetime of devices~\cite{adu2018energy} and allowing them to operate in remote areas without easy access to mains electricity to create Internet of remote things~\cite{9222142} or ocean of things~\cite{9086780}.
Apart from these technologies, sensor devices can also be recharged by a robot/drone, see \cite{8869810, 9468714} for examples.
%

\begin{table*}[htbp]
\caption{Ambient energy sources.}\label{table:ehsource}
\footnotesize
\newcommand{\tabincell}[2]{\begin{tabular}{@{}#1@{}}#2\end{tabular}}
\centering
\begin{tabular}{|l|l|c|c|c|c|}
\hline
\textbf{Energy Source} & \textbf{Work} & \tabincell{l}{\textbf{Energy} \\ \textbf{Rate/Density}} & \tabincell{l}{\textbf{Conversion} \\\textbf{Efficiency}} & \tabincell{l}{\tabincell{c}{\textbf{Harvested} \\ \textbf{Energy}}} & \tabincell{l}{\textbf{Output}\\ \textbf{Voltage}}\\
\hline
\hline
\multirow{2}{*}{ {\em Solar}} & ~\cite{hamza2021hybrid} & 100 mW/cm$^2$ & 16-17\% & 192.9 $\mu$W & 6.3 V\\
\cline{2-6}
& ~\cite{sharma2018an} & 1000 W/cm$^2$ & 15\% & 1.8 mW &  6 V\\
\hline
\multirow{2}{*}{ {\em Wind}} & ~\cite{tan2011optimized} & 82 mW at 3.36 m/s & 39\% & 3.6 mW & 4 V\\
\cline{2-6}
& ~\cite{haidar2019an} & 15 mW at 4 m/s & 52.86\% & 7.6 mW & 0.8 V\\
\hline
 {\em Piezoelectric} & ~\cite{oh2017powering} & 50 $\mu$J/N & 46\% & 11.1 $\mu$W & 0.7 V \\
\hline
{\em Kinetic} & ~\cite{magno2018micro} & N/A & N/A & 624 $\mu$W & N/A \\
\hline
{\em Thermal} & ~\cite{leonov2013ther} & 25-60 mW/cm$^2$ & N/A & 50 $\mu$W & N/A \\
\hline
\end{tabular}
\end{table*}

In this paper, we focus on Radio Frequency (RF) energy harvesting IoT networks.
Briefly, these networks take advantage of radio signals emitted in frequency range from 3 kHz to 300 GHz~\cite{lu2015wireless}. The amount of harvested energy is a function of i) an RF source's transmit power, ii) operational frequency, and iii) the distance between an RF source and sensor devices. Table~\ref{table:rfsource} compares various RF sources as per the aforementioned factors. 
RF sources can mainly be classified into two types: {\em ambient} and {\em dedicated}. 
Examples of ambient RF sources include television tower, WiFi, and base stations operating in 4G/5G systems.  A prototype device that relies on an ambient source is reported in~\cite{vyas2012battery}, where the device is able to harvest  70 mW of RF energy within 3 minute from a television tower at a distance of 6.5 km.  On the other hand, the prototype in~\cite{talla2015powering} has a camera and harvests RF signals in a WiFi network.  It harvests energy at a rate of 2.77 $\mu$J/s at a distance of up to 8.53 meter. 
%

As for dedicated RF sources, they are either co-located with an Access Point (AP), so called Hybrid Access Point (HAP), or deployed strategically for the purpose of charging devices; these sources are also called power beacons or stations.
In this respect, prior works such as~\cite{Stoopman2013a} have demonstrated  a dedicated RF transmitter that provides a maximum transmit power of 1.78 W at 868 MHz. Sensor devices are able to harvest at least -26.3 dBm at a distance of 25 meter. Further, there are now commercial RF transmitters, e.g., RF transmitter TX91501 sold by Powercast~\cite{powercast}.

\begin{table*}
\caption{Example RF energy sources.}\label{table:rfsource}
\footnotesize
\newcommand{\tabincell}[2]{\begin{tabular}{@{}#1@{}}#2\end{tabular}}
\centering
\begin{tabular}{|c|c|l|c|c|c|c|c|c|c|}
\hline
\tabincell{c}{\textbf{ Source Type} }& \tabincell{c}{\textbf{Energy} \\\textbf{Source}} & \tabincell{c}{\textbf{Work}} & \tabincell{c}{\textbf{Transmit Power/} \\\textbf{\textbf{ Energy Density}}} & \tabincell{c}{\textbf{Conversion} \\ \textbf{Efficiency}} & \tabincell{c}{\textbf{Harvested} \\ \textbf{Energy}} & \tabincell{c}{\textbf{Output}\\ \textbf{Voltage} }& \tabincell{c}{\textbf{Distance}} & \textbf{Sensitivity} & \tabincell{c}{\textbf{Frequency} }\\
\hline
\hline
\multirow{5}{*}{ {\em Ambient RF}} & \tabincell{c}{Television \\tower}  & ~\cite{vyas2012battery} & 10-100 kW & 70 mW & N/A &  3 V & 6.5 km & N/A & 550 MHz  \\
\cline{2-10}
 & \tabincell{c}{Digital \\ television} & ~\cite{sidhu2019a} & 960 kW & N/A & 60 $\mu$W & 5V  & 4 km  & N/A  &  500 MHz\\
 \cline{2-10}
& \multirow{2}{*}{WiFi} & ~\cite{sidhu2019a} & 0.1 $\mu$W/cm$^2$ & 33.7\% & 76.35 $\mu$W & N/A & N/A & N/A & 2.4 GHz \\

\cline{3-10}
& & ~\cite{talla2015powering} & N/A & N/A & 2.77 $\mu$J/s & 2.3/2.4 V & 8.53 m & N/A & 2.4 GHz \\

 \cline{2-10}
& GSM & ~\cite{vi2013rf} & 0.1-3.0 mW/cm$^2$ & N/A & 100 $\mu$W & N/A & 20-100 m & N/A & 900/1800 MHz\\
\hline
\multirow{6}{*}{{\em Dedicated RF}} & \multirow{5}{*}{ \tabincell{c}{Isotropic RF\\ transmitter} } & ~\cite{Stoopman2013a} & 1.78 W  & 31.5\%  & N/A & 1 V & 25 m & -26.3 dBm &  868 MHz  \\
\cline{3-10}
& & ~\cite{hammi2008a} & 100 W & N/A &  1 $\mu$W & N/A & N/A & N/A & 1960 MHz \\
\cline{3-10}
& & ~\cite{st2014co} & 1.78 W  & 40\%  & N/A & 1 V & 27 m & -27 dBm &  868 MHz  \\
\cline{3-10}
& & ~\cite{agr2016design} & 10 mW  & N/A & 0.3 mW  &  1.5 V  & N/A  &  -20 dBm  & 900 MHz \\
\cline{3-10}
& & ~\cite{hu2017a} &  N/A & 30.1\% & N/A   &  30 $\mu$W  & 1.2 V &  N/A & 900 MHz \\
\cline{2-10}
& \tabincell{c}{TX91501 \\Powercast \\ transmitter} & ~\cite{powercast} & 4 W  & 62\%  & 1 $\mu$W & N/A &  11 m &  -12 dBm &  915 MHz \\
\hline
\end{tabular}
\end{table*}

%

\begin{figure}[ht]
    \centering
    \includegraphics[width=0.5\textwidth]{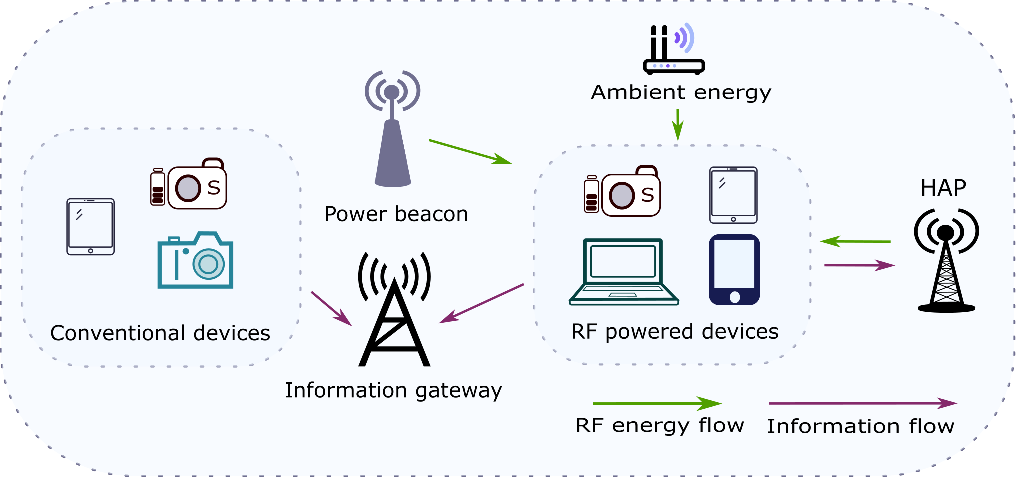}
    \caption{An example IoT network with RF powered and conventional devices.}
    \label{fig:rf-iot}
\end{figure}

Figure~\ref{fig:rf-iot} shows an RF energy harvesting IoT network.
In general, such a network has three components: one or more RF energy transmitters, information gateways and sensing devices. 
These devices can either receive energy from ambient or dedicated RF sources. The information gateways collect data from devices, and may have a renewable energy source. Devices harvest RF energy and upload data to an information gateway. In some works, see~\cite{lu2015wireless, 7539589}, the information gateway is also called an RF energy transmitter or a HAP. 
%

Medium Access Control (MAC) or link scheduling protocols play a critical role in wireless networks~\cite{9434395} where nodes share the same spectrum or frequency bands. 
These protocols ensure transmissions or links do not interfere with one another; equivalently, they ensure the Signal-to-Noise plus Interference (SINR) ratio at receivers does not exceed a predefined threshold.  
For example, in an RF-energy harvesting IoT network, devices will have to transmit their sensed data to an HAP.  Consequently, depending on the decoding capability of the HAP, it may be capable of decoding a single transmission at a time only.  Alternatively, if it supports Successive Interference Cancellation (SIC)~\cite{islam2017power}, then it is able to decode concurrent transmissions, subject to each transmission satisfying an SINR condition~\cite{8972552}.  Further, an HAP or power beacons may use the same channel to deliver energy or/and data to devices.  Hence, a key concern is that the channel process does not degrade the throughput of non-energy harvesting devices significantly; i.e., the issue concerns the co-existence of charging and data transmissions, see Section~\ref{sec:CSMA} for more details.

\begin{table*}
\caption{A comparison to prior surveys.  Each tick indicates 10 papers reviewed by a reference (rounded up). }\label{table: compare}
\footnotesize
\newcommand{\tabincell}[2]{\begin{tabular}{@{}#1@{}}#2\end{tabular}}
\centering
\begin{tabular}{|c|c|c|c|c|c|c|c|}
\hline
\multirow{3}{*}{Ref.} & \multirow{3}{*}{Dedicated RF} & \multirow{3}{*}{Ambient RF} &  \multicolumn{5}{c|}{ {Medium Access Control Protocols}}  \\ 
\cline{4-8}
  &   &   &  \multicolumn{2}{c|}{ { Contention-Based}} & \multicolumn{3}{c|}{ { Contention-Free}}  \\ 
\cline{4-8}
  &   &   & Aloha & CSMA & Polling & Dynamic TDMA & NOMA \\ 
\cline{4-8}
\hline
\hline
\cite{sherazi2018comprehensive} & \Circle  & \CIRCLE & - & \checkmark \checkmark & \checkmark &  - & -  \\
\hline
\cite{kosunalp2015mac} & \Circle  & \CIRCLE & \checkmark & \checkmark & \checkmark & -  & -  \\
\hline
\cite{Quintero2019improvements} & \CIRCLE  & \CIRCLE & - & \checkmark & \checkmark & - & - \\
\hline
\cite{Parisa2015Overview} & \Circle  & \CIRCLE & - & \checkmark & \checkmark & - & -  \\
\hline
\cite{kaur2019recent} & \Circle  & \CIRCLE & \checkmark & \checkmark & \checkmark & \checkmark & - \\
\hline
\cite{Famitafreshi2021comprehensive} & \Circle  & \CIRCLE & \checkmark & \checkmark \checkmark & \checkmark & \checkmark \checkmark & \checkmark \\
\hline
\cite{Sherazi2022comprehensive} & \CIRCLE & \Circle & - & \checkmark & - & \checkmark & - \\
\hline
Ours & \CIRCLE & \Circle & \checkmark \checkmark  & \checkmark  \checkmark & \checkmark \checkmark \checkmark \checkmark & \checkmark \checkmark  & \checkmark \checkmark   \\
\hline
\end{tabular}
\end{table*}

%
%
To date, there are no comprehensive surveys that focus on MAC or channel access protocols that are designed to operate in IoT networks with one or more {\em dedicated} RF energy sources.
This paper fills this critical gap in the literature.
Table~\ref{table: compare} shows related surveys that have considered energy harvesting channel access protocols. we see that our survey has a number of differences. The main difference is that we focus on MAC protocols with dedicated RF energy sources. 
In this respect, only the authors of~\cite{Quintero2019improvements} and \cite{Sherazi2022comprehensive} have considered dedicated RF sources. However, they only reviewed a handful of works. The work in \cite{sherazi2018comprehensive} and \cite{Famitafreshi2021comprehensive} has reviewed MAC protocols that employ Aloha, CSMA, polling, dynamic TDMA and NOMA.  However, their focus is on devices with an ambient RF energy source.   Lastly, works such as~\cite{kosunalp2015mac, Parisa2015Overview} and \cite{kaur2019recent} only briefly discussed contention-based and/or contention-free MAC protocols that use ambient RF energy. 
In summary, this paper makes the following contributions:
\begin{itemize}
    \item It reviews works on channel access protocols designed for IoT networks with one or more {\em dedicated} energy sources.  
    \item It highlights the main network architectures and time frame/slot structure used for charging or energy delivery by HAP(s), and data transmissions from devices.
    \item It covers contention-based and contention-free protocols, and highlights the key parameters that are optimized by these protocols, and also requirements such as minimum data rate.    
    For each category of protocols, we qualitatively compare the works that adopt the same protocol.   We also compare both categories of protocols, highlight their research aim and key ideas.
    \item It presents an extensive list of future works, which include state-of-the-art technologies such as intelligent reflective surfaces~\cite{9326394}, predict-and-optimize framework~\cite{SPO} and graphical neural networks~\cite{9046288}.
\end{itemize}
%

We {\em emphasize} that the following topics/areas are out of scope.  First, this survey {\em does not} consider works that assume ambient energy sources; e.g., devices operating in a cognitive radio network or WiFi networks whereby an RF emitter is not aware of RF-energy harvesting devices.
Second, it excludes works whereby devices have a fixed transmission order or time.  For example, the majority of wireless powered communication network (WPCN) works, see \cite{lu2015wireless}, assume devices are pre-assigned to a time slot or have a given transmission order, e.g., based on their channel gain~\cite{nasir2019noma}.  In other words, we only include works that have the sub-problem of determining the transmission time/order of devices or/and energy transmitters.
Third, this survey omits works that charge devices wirelessly using one or more mobile nodes, e.g., a drone~\cite{8422707, 8839870}.  This is because their focus is on trajectory planning.  
%

Next, i.e., Section~\ref{SECBG}, we first present key network architectures and channel access protocols employed in prior works. Further, we outline time structures used for charging devices and data transmissions.  We also briefly explain the main operating principles of channel access protocols used in prior works.
After that, in Section~\ref{sec:cont-based} and \ref{CFbasedProtocols}, we respectively outline two categories of channel access works, namely contention-based, and contention-free channel access.
%

%

%
%

%
\section{Background}\label{SECBG}
\subsection{Network Architectures}
\begin{figure*}[tp]
\centering
\subfloat[IoT network with an HAP.] 
{\includegraphics[width=.29\textwidth]{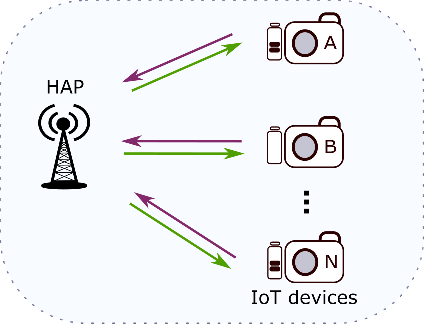}\label{fig:iotnetwork1}}\hspace*{-0.0em}
\hfill
\subfloat[IoT network with power beacons.]
{\includegraphics [width=.39\textwidth]{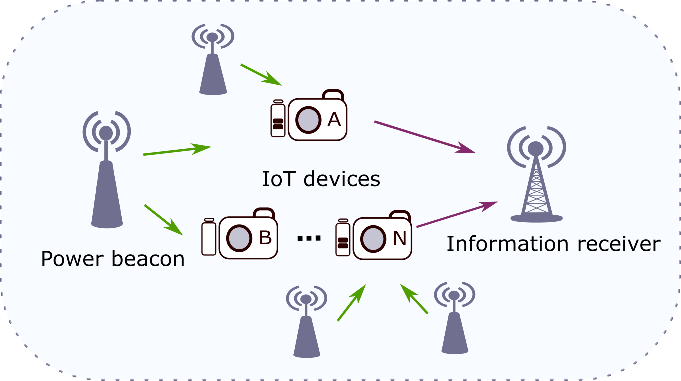}\label{fig:iotnetwork2}}\hspace*{-0.0em}
\hfill
\subfloat[IoT network with relays.]
{\includegraphics [width=.32\textwidth]{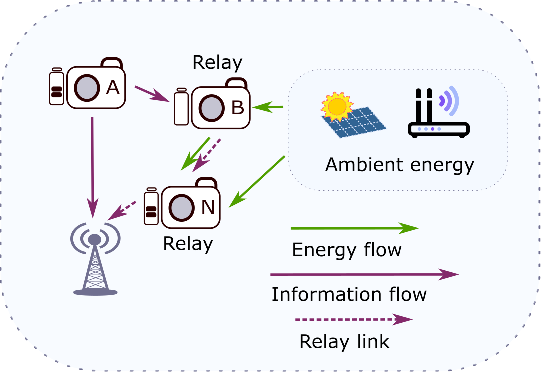}\label{fig:iotnetwork3}}
\caption{Typical IoT networks: (a) an HAP charges device(s).   Devices then use their harvested energy and reply to the HAP using a channel access method, (b) devices are charged by one or more power beacons. This architecture helps overcome the doubly near-far problem, and (c) some devices are used as relays to help forward data.  Further, these relays may harvest energy from ambient or dedicated source(s).}
\label{fig:iotnetworks}
\end{figure*}

Figure~\ref{fig:iotnetworks} shows typical systems considered by prior works.  Specifically, an IoT network has one or more energy transmitters.  Further, these energy transmitters can be equipped with multiple antennas to enable beamforming~\cite{yang2015throughput, boshkovska2017robust}, full-duplex communications~\cite{asiedu2020beamforming} or SIC~\cite{li2021random, 9994069}.  Moreover, devices may have multiple antennas, e.g., \cite{Rubio2019}.   
In general, energy and data transmissions use the same frequency band.  However, some works, e.g.,~\cite{9084128}, have considered frequency/channel assignment.
%
Figure~\ref{fig:iotnetwork1} shows a popular setup.  The HAP is responsible for charging devices and data collection.  
In Figure~\ref{fig:iotnetwork2}, one or more power beacons are responsible for charging devices, which then use their harvested energy to transmit data to an information receiver.  The main reason for deploying power beacons is to address the doubly near-far problem~\cite{ju2013throughput}, whereby a device that is far away from a HAP has a low energy harvesting rate as well as a low transmit power or data rate.  In this respect, a power beacon can be placed near such a device, and thereby improve its energy harvesting rate and transmit power. 
Further, these power beacons may cooperatively charge devices~\cite{he2020maximizing, 7945489, naderi2014frmac}.
Figure~\ref{fig:iotnetwork3} shows a setting with one or more relays.  They help forward data/signal to a receiver and also act as a charger~\cite{9145865, 8049272}.  Further, they may be capable of harvesting ambient energy, see for example the relay in~\cite{9615376}.  
%

\begin{figure*}[tp]
\centering
\subfloat[A time switching receiver.] 
{\includegraphics[width=.47\linewidth]{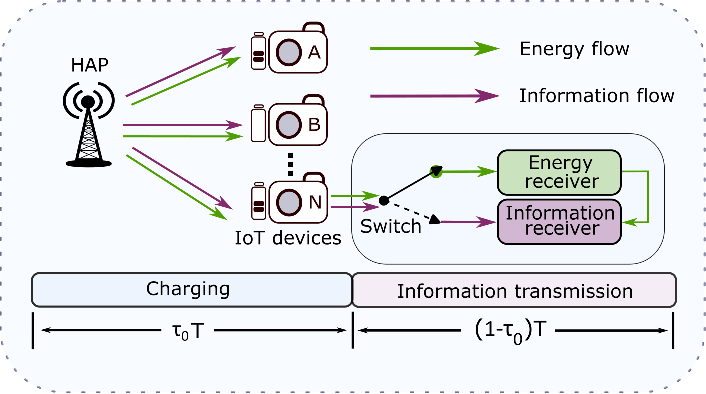}\label{fig:ts}}\hspace*{-0.0em}
\hfill
\subfloat[A power splitting receiver.]
{\includegraphics [width=.47\linewidth]{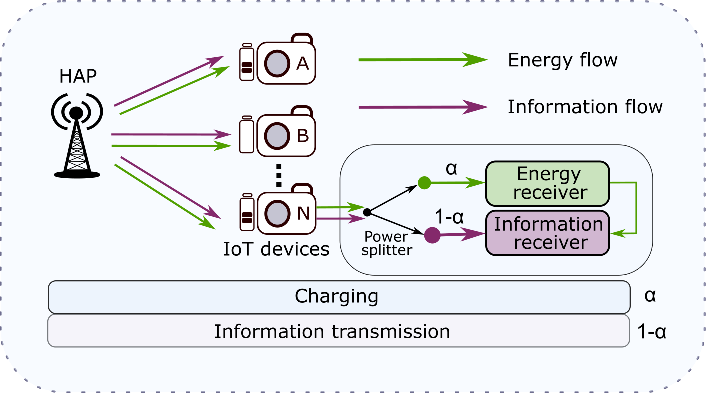}\label{fig:ps}}\hspace*{-0.0em}
\caption{Wireless powered or RF charging IoT networks, and receiver architectures.}
\label{fig:recearch}
\end{figure*} 
Figure~\ref{fig:recearch} shows two common receiver architectures that realize simultaneous wireless information and power transfer (SWIPT).
In Figure~\ref{fig:ts}, SWIPT divides a frame into two parts.  An HAP or power beacon first charges devices at a given transmit power for $\tau_0T\in [0,1]$ seconds.  After that, the remaining time is used to receive data, where devices transmit to a receiver.  Here, the transmission order is either fixed or decided by some channel access method.  

Another SWIPT architecture is depicted in Figure~\ref{fig:ps}.  A fraction, i.e., $\alpha\in [0,1]$, of the received power is diverted to an energy harvester whilst the remaining signal power, i.e., $1-\alpha$, is sent to the information decoder.  
Observe that $\alpha$ controls the trade-off between the amount of energy harvested by a device and its information rate.  This will be an issue when an HAP aims to maximize both uplink and downlink rate.

\begin{figure}[ht]
    \centering
    \includegraphics[width=0.5\textwidth]{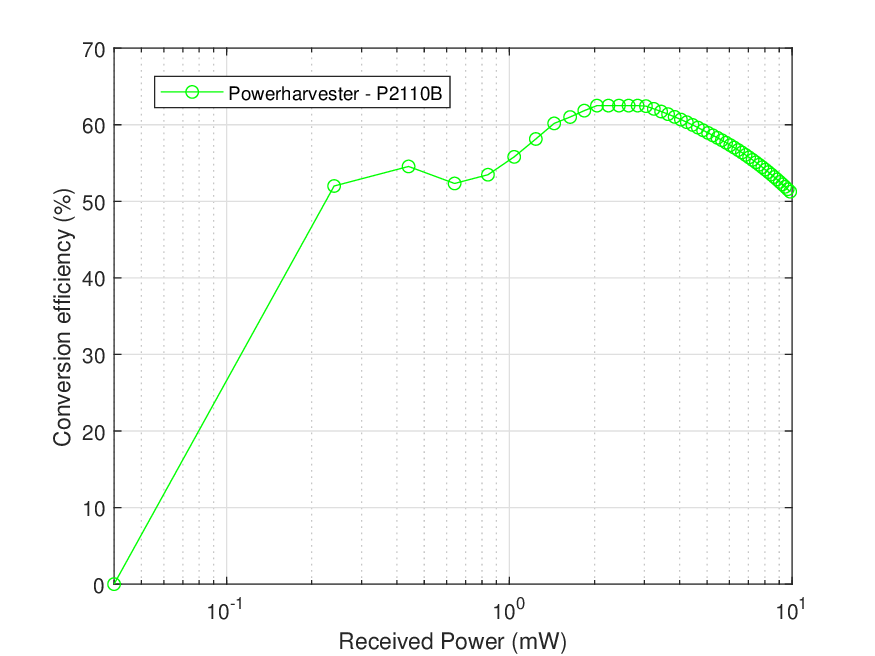}
    \caption{Energy conversion efficiency of the P2110B RF power harvester~\cite{powercast}.}
    \label{fig:ce2110}
\end{figure}

We also point out that RF-energy harvesting efficiency is non-linear, and it is a function of the received power~\cite{boshkovska2015practical}.
Figure~\ref{fig:ce2110} shows an example model.  In this respect, a key issue is ensuring an HAP uses a transmit power that yields the highest energy conversion efficiency.

%

%

%
\begin{figure*}[tp]
\centering
\subfloat[Harvest-then-transmit.] 
{\includegraphics[width=.333\textwidth]{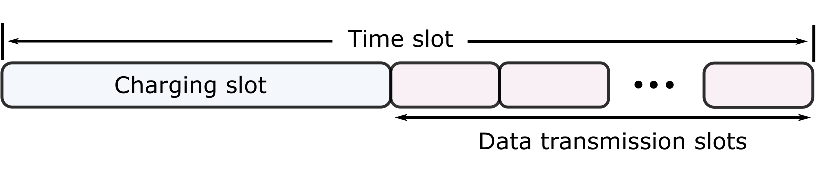}\label{fig:timeslot1}}\hspace*{-0.0em}
\hfill
\subfloat[Multiple time slots.]
{\includegraphics [width=.3\textwidth]{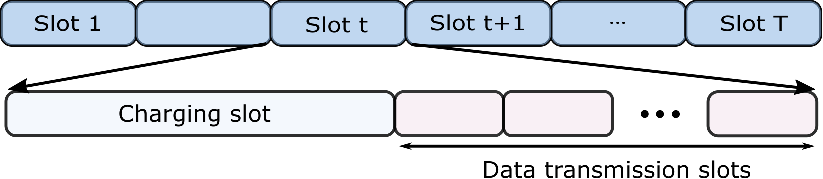}\label{fig:timeslot2}}\hspace*{-0.0em}
\hfill
\subfloat[Time slots with mode selection.]
{\includegraphics [width=.333\textwidth]{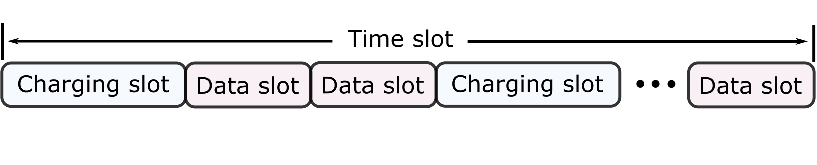}\label{fig:timeslot3}}
\caption{Time slot structures used by prior works.}
\label{fig:timeslot}
\end{figure*} 
%

\subsection{Time Slots Structure}
Time is a key resource in prior works, where nodes in IoT networks have to decide how to share a wireless channel to carry out energy delivery or/and data transmissions. 
To do that, time is divided into discrete time slots.  In this respect, a standard assumption is that the channel condition of devices in each time slot or frame is static; i.e., prior works generally assume block fading.

A popular structure is time switching or harvest-then-transmit~\cite{ju2013throughput, lu2015wireless}; see Figure~\ref{fig:timeslot1}.  Note that the majority of prior works assume devices are pre-allocated a time slot.  Specifically, each device knows exactly its transmission time slot, meaning channel access is not an issue.   
On the other hand, for works that use random access, the number of time slots or frame size can be adjusted and devices select a time slot based on a probability.
Another time structure contains multiple time slots; see Figure~\ref{fig:timeslot2}.  
In this respect, Figure~\ref{fig:timeslot3} shows another variant of the multiple time slots structure, whereby each time slot has one of two modes: (i) charge, or (ii) data.  
A HAP uses a time slot for charging if its channel gain to devices is favourable.  
Otherwise, a time slot is used for data upload; see \cite{10109156, song2021novel} for more details.
The main challenge is that a node has non-causal knowledge of channel conditions, which affect its available energy in future time slots.  Further, a device may not be aware of its future traffic demand.  

Another issue to consider in the multiple time slots case is the proportion of downlink and uplink transmissions; i.e., more downlink transmissions result in users harvesting more energy, which lead to higher transmit power; equivalently, higher uplink transmission rates.  On the other hand, more downlink transmissions reduce uplink transmission opportunities.  Hence, there is a trade-off between uplink and downlink transmission duration.   Further, there is the issue of fairness between total uplink and downlink rate, or the uplink or downlink of individual users.
\subsection{Uplinks and/or Downlinks}\label{NOMAULDL}
Most works aim to charge devices in downlink transmissions in order to maximize the amount of data transmitted by devices to an information receiver or HAP.
To do this, the main resources to optimize include (a) time used to charge devices, (b) transmit power of a HAP or power beacon, (c) number of energy transmitters, (d) transmit power of devices, (e) uplink slot size, (f) channel, and/or (g) group or pair of transmitting devices, to name a few.
A key consideration is the channel condition to/from devices, or/and between a relay and devices/HAP; this is in addition to their energy level.
A challenge is that an HAP may have imperfect knowledge of the channel condition to devices and their traffic arrivals.
Note that it is impractical to collect channel gain or/and energy level information in large scale IoT networks.  Further, collecting such information requires devices to spend their precious harvested energy~\cite{9783143, 9739685}.

Another challenge is the doubly near-far problem, whereby a device located far away from an RF energy source will experience a poor channel gain.  Consequently, it has a low amount of RF-energy, and thus it can only transmit at a low transmit power.  Further, as the device may be far from its receiver, its channel gain will be poor.  These factors mean the device will have a low data rate.  In this respect, a key goal is to ensure all devices have a fair throughput or to optimize the worst data rate~\cite{ju2013throughput}.
Some works aim to deliver data to devices.    Specifically, an HAP sends a poll for data along with a message to a device, see e.g., see Section~\ref{POLLDL} for details.  In this respect, how often a device is polled affects its downlink throughput.  Further, a key issue is ensuring devices have sufficient energy to receive the poll and message.
Alternatively, a multi-antenna HAP may use NOMA to deliver data to multiple devices concurrently, see \cite{8485639}.

Lastly, a small number of works aim to optimize both uplink and downlink sum-rates~\cite{7998247} or to consider throughput fairness of both uplinks and downlinks~\cite{8294215}.
The problem is challenging as a HAP has to deliver both energy and data in downlink transmissions~\cite{7998247}.
Specifically, receivers may employ a power split receiver architecture, where devices have to optimize their power split ratio that determines the amount of harvested energy versus information rate~\cite{liu2013wireless}.  In this respect, there is a trade-off between downlink and uplink rate, where a device that prefers to harvest more energy will sacrifice its downlink rate, and vice-versa.
This trade-off will be discussed in Section~\ref{POLLUPDOWN} and \ref{NOMAULDL}.

%
\begin{figure*}[tp]
\centering
\subfloat[IRSA.] 
{\includegraphics[width=.5\textwidth]{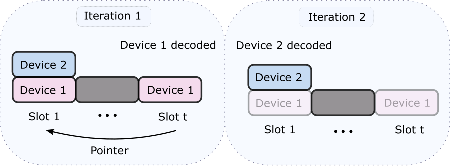}\label{fig:iras}}\hspace*{-0.0em}
\hfill
\subfloat[NOMA.]
{\includegraphics [width=.5\textwidth]{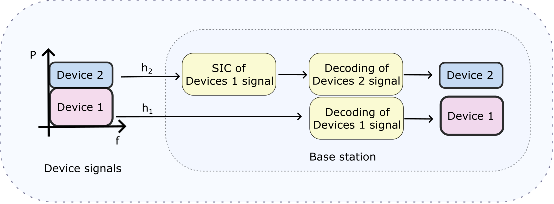}\label{fig:noma}}\hspace*{-0.0em}
\caption{Two example MAC protocols.}
\label{fig:MAC}
\end{figure*} 
\subsection{Channel Access}
To date, prior works have studied/proposed the following MAC protocols: 
\begin{enumerate}
\item {\em Aloha}, which includes Slotted Aloha, framed slotted Aloha and irregular slotted Aloha (IRSA)~\cite{ghanbarinejad2013irregular}.  
Note that Aloha based protocols are widely used for reading RFID tags, see \cite{ChinRFIDSurvey}.
In Slotted Aloha, devices use a probability to determine whether to transmit.  
As for IRSA, devices send multiple copies of a packet to a receiver, which then uses the said copies to decode packets iteratively; In see Figure~\ref{fig:iras}, Device-1 retransmits a copy of the packet it transmitted in slot-1. A receiver can then use it to decode the packet from Device-2.
\item {\em Carrier Sense  Multiple Access with Collision Avoidance (CSMA/CA)}, where a device first senses the channel and transmits if the channel is free.  Otherwise, it carries out a backoff process.  Further, a device freezes its backoff counter if it experiences an energy outage.  Apart from that, a Request-to-Send (RTS) and  Clear-to-Send (CTS) message can be used to indicate or reserve a channel for charging purposes.
\item {\em Time Division Multiple Access (TDMA)}, where the size of each time slot is optimized according to the transmit power or/and channel gain to/from a device.  A variant of TDMA is {\em dynamic} TDMA, whereby devices are {\em not} pre-assigned to any time slot(s).  Hence, a key issue is scheduling one or more devices into a time slot or to derive a link schedule~\cite{8972552}.
\item {\em Non-Orthogonal multiple access (NOMA)}~\cite{9831440}.  These methods allow one-to-many, and many-to-one communications.  Consequently, NOMA/SIC helps avoid collision and have a higher spectral efficiency.  Figure~\ref{fig:noma} provides the key concepts of NOMA.  An important issue is selecting or grouping devices and setting their transmit power to ensure decoding success.

\item {\em Polling}, where a gateway selects $K$ devices based on some strategy, e.g., select devices with the most energy.  Its goal is to meet some objectives, e.g., average throughput, fairness or Age of Information (AoI).  Further, the gateway may piggyback a poll message in a packet destined for selected devices.
\end{enumerate}
Next, Section~\ref{sec:cont-based} discusses contention-based protocols, namely Aloha and CSMA.  We highlight how prior Aloha-based works minimize collision and consider the energy availability of devices.  Then we highlight CSMA works, which in addition to data transmissions, devices also use CSMA to send a charging request message to one or more power beacons.
Section~\ref{CFbasedProtocols} discusses contention-free protocols, namely polling, TDMA and NOMA.   

\section{Contention-based channel access}\label{sec:cont-based}
Random access protocols have many advantages.  A HAP does not have to pre-allocate time slot(s) to devices, meaning devices contend for a channel in an on-demand manner and in a distributed manner.  Further, they do not have to assume devices have energy or/and data whenever their allocated time slot arises.  
In practice, determining the energy level and/or channel gains require excessive signaling, which involves sending out pilot symbols or request to each device.  This is not practical especially in large-scale IoT networks~\cite{9783143}. 
Hence, random channel access protocols are well-suited for use in such networks whereby devices transmit whenever they have sufficient energy.
Another issue is transmission delay.  Specifically, in large-scale networks, if devices are allocated a time slot, they may have to wait a significant amount of time before they can transmit their data.  
Lastly, IoT devices may transmit small packets, meaning it is not reasonable for them to first reserve channel resources, which incurs non-negligible signaling cost~\cite{choi2019harvestor}. 

Next, we discuss works that employ Slotted Aloha and CSMA.  
%
%
As it will become clear in our discussion, these works aim to optimize one or more of the following variables: 
(i) number of time slots in each frame or frame size, (ii) transmission probability of devices, and (iii) number of replicas of each packet used by each device.
Apart from that, some works have considered NOMA, e.g.,~\cite{li2021random}, whereby a receiver (or HAP) decodes concurrent transmissions using SIC.
As for CSMA, prior works propose CSMA/CA based protocols for use by energy transmitters and devices.   In particular, energy transmitters use CSMA/CA before sending out charging packets or to reserve the channel for energy delivery.  On the other hand, devices use CSMA/CA to send a charging request.
%
%

\subsection{Slotted Aloha}
We start with works that aim to optimize the frame size used by devices.
Choi et al.~\cite{choi2018slotted, choi2019harvestuntil} aim to maximize throughput and fairness in an IoT network where an HAP has a full-duplex radio.  Further, there are low and high priority devices.   In particular, low priority devices, i.e., those farther away from the HAP, access the channel at a later time; doing so allows these devices to spend more time to harvest energy.  The authors then derive the optimal number of slots or frame size that maximizes throughput.  
The same authors also consider a harvest-or-access slotted Aloha protocol~\cite{choi2019harvestor}, where an HAP opportunistically performs wireless energy transfer to devices during idle random access slots. They also derived the optimal number of slots or frame size that leads to the highest throughput; the said slots are used by the HAP to transfer energy, and data transmissions.
On the other hand, Yu et al.~\cite{yu2019two} consider a dynamic frame slotted Aloha method.  Devices are not aware of the energy level of other devices.  The HAP in \cite{yu2019two} aims to maximize system throughput using imperfect imperfect battery level of devices/users.  It adopts a two-layer approach, where at the first layer, it optimizes its transmit power and frame size.   Given a frame size, at the second layer, users then optimize their slot selection.   
%

Some works set the optimal frame size using reinforcement learning~\cite{Sutton}.
Note, works that apply reinforcement learning in {\em non-energy} harvesting networks, e.g.,~\cite{chu2012reinforcement,zhang2022making}, are beyond the scope of this paper. 
%
To date, there are only three reinforcement learning works that have considered dedicated energy transmitters.
These works use reinforcement learning to learn the optimal frame size or number of slots for energy harvesting and data transmissions.  Note, an agent or HAP adapts these quantities over time based on the number of devices that have sufficient energy to transmit.
Specifically, the HAP in~\cite{iqbal2019learning} adapts the frame size used by devices using Q-learning~\cite{watkins1992q}.  Its goal is learn the optimal frame size that minimizes idle and collision slots. 
In a different work, Li et al.~\cite{li2021random} consider a SIC-capable HAP.  Li et al. employ Q-learning to determine the optimal transmit power of each device by only observing the battery level of each device without knowledge of CSI in order to maximize system sum-rate.    Advantageously, their solution is distributed whereby devices are able to learn the optimal policy to set their transmit power that facilitates SIC decoding.
Lastly, the HAP and devices in~\cite{9632809} use IRSA to access the channel.  Further, based on their energy state, devices then use Q-learning to determine a policy that governs the number of replicas used for each packet transmission.
Collision is a key issue as it reduces throughput and critically, and wastes the energy harvested by devices.
Hence, a number of works have considered different methods to reduce collisions.
For example, the HAP in~\cite{silva2021slotted} controls its frequency and duration of charging.  Specifically, the frequency of charging is determined by a threshold, whereby once the number of transmission attempts exceed the threshold, the HAP sends a wake-up signal.  Its charging duration (in terms of slots) can also be optimized, whereby the HAP informs devices the number of time slots used to send energy packets.  After receiving these energy packets, devices access the channel using slotted Aloha.  In this respect, the authors also introduce two operational modes: hold-before-charge (HBC) and drop-before-charge (DBC).  These modes govern when a device transmits a packet that arrives in the same time slot when a wake-up signal is transmitted.  In HBC, devices the packet immediately after receiving all energy packets.  As for DBC, devices drop the packet.
On the other hand, the work in~\cite{deng2017towards} considers the fact that devices are charged at a different rate and thus transmit at different times.  Hence, a key problem is to determine the optimal transmit power used to charge devices.  Critically, if devices attain sufficient energy at different times, then collision is minimized or eliminated. 

Another way to minimize collision is to leverage NOMA.  Specifically, a receiver or HAP has SIC capability, and thereby allowing it to decode concurrent transmissions, assuming some conditions are met~\cite{li2021random, tegos2020slotted}.   
In this respect, the work in~\cite{tegos2020slotted} considers a NOMA-based WPCN, where a HAP first charges devices.  This is then followed by an information phase, where each user has a transmission probability.  Further, the authors of~\cite{tegos2020slotted} consider two multi-user detection techniques to resolve collisions, namely SIC and joint decoding\footnote{The main difference between joint decoding and SIC is that the former does not involve subtraction of signals or iterative decoding.}. 
In terms of fairness, the work in~\cite{8464255} sets a channel access probability based on the distance of each device. The work in \cite{8464255} further considers HAP transmit power, energy harvesting duration, and the transmission rate of each device.
The devices/users in~\cite{pejoski2019rf} only transmit when their channel is favorable.  During poor channel condition, they conserve energy. Moreover, in~\cite{pejoski2019rf},  each user must ensure a given minimum received power.
Pejoski et al.~\cite{pejoski2020slotted} consider Nakagami-m fading channels; this is in contrast to \cite{pejoski2019rf}, where the authors assume Rayleigh fading and their HAP uses two possible schemes.   
The first scheme uses a fixed rate, which aims to meet a given received power at an HAP. The second scheme uses variable rates, whereby the HAP specifies different data rates.  In both schemes, a user must optimize its transmit power to meet a specified data rate.
Lastly, in IoT networks with half-duplex devices, they may miss their recharging opportunity when they transmit data.
%
To this end, the proposed protocol in~\cite{lin2020framed} ensures that the charging time slot selected by a full-duplex HAP to charge a half-duplex device is not the same as the data transmission slot chosen by the said device; note, devices run slotted Aloha.  Advantageously, it avoids wasting charging opportunity by focusing only on devices that are able to receive a charge.
%
\subsubsection{Discussion}
Referring to Table \ref{Table aloha}, the majority of works employ framed slotted Aloha.  Only a few works, namely \cite{yu2019two, li2021random, iqbal2019learning}, have considered adapting the frame size to account for time varying energy harvesting rate at devices.  Besides frame size, prior works have adapted the charging policy or frequency of a HAP. Specifically, the work in \cite{choi2019harvestor} takes advantage of idle slots, and the work in \cite{silva2021slotted} uses the number of transmissions to ascertain whether devices requiring charging.
Lastly, an effective method to deal with collisions is to employ SIC, see \cite{li2021random, tegos2020slotted}.

\begin{table*}
\Large
\newcommand{\tabincell}[2]{\begin{tabular}{@{}#1@{}}#2\end{tabular}}
\caption{A comparison of Aloha works.}\label{Table aloha}
  \centering
  \resizebox{\textwidth}{!}{
  \begin{tabular}{|c|c|c|c|l|l| }
\hline
\textbf{Energy Source}&\textbf{Works}&\textbf{Uplink Channel Access}&\textbf{Performance Objective(s)}&\textbf{Key Idea(s)}\\ [0.5ex]
\hline
\hline
\multirow{18}{*}{HAP} &
\textbf{\cite{choi2018slotted}}&\tabincell{l}{Framed Slotted Aloha}&\tabincell{l}{Throughput,\\Fairness}&\tabincell{l}{Harvest until access, transmission priority \\based on device distance}\\ 
\cline{2-5}&
\textbf{\cite{choi2019harvestuntil}}&\tabincell{c}{Framed Slotted Aloha}&\tabincell{l}{Throughput}&\tabincell{l}{Harvest until access}\\ 
\cline{2-5}&
\textbf{\cite{choi2019harvestor}}&\tabincell{c}{Slotted Aloha} &\tabincell{l}{Throughput}&\tabincell{l}{Take advantage of idle slots for charging}\\ 
\cline{2-5}&
\textbf{\cite{yu2019two}}&\tabincell{c}{Dynamic framed\\Slotted Aloha}&\tabincell{l}{Throughput}&\tabincell{l}{Adaptive HAP transmit power, frame size,\\ and slot selection}\\ 
\cline{2-5}&  
\textbf{\cite{li2021random}}&\tabincell{c}{Framed Slotted Aloha}&\tabincell{l}{Throughput}&\tabincell{l}{SIC and adaptive transmit power control}\\
\cline{2-5}&
\textbf{\cite{9632809}}&\tabincell{c}{IRSA}&\tabincell{l}{Throughput}&\tabincell{l}{Optimize the number of replicas}\\  
\cline{2-5} 
%
& \textbf{\cite{silva2021slotted}}&\tabincell{c}{Slotted Aloha}&\tabincell{l}{Throughput}&\tabincell{l}{Set charging time or frequency as per\\number of transmissions}\\ 
\cline{2-5}&
\textbf{\cite{deng2017towards}}&\tabincell{c}{Slotted Aloha}&\tabincell{l}{Collisions}&\tabincell{l}{Optimize HAP transmit power}\\ 
\cline{2-5}&
\textbf{\cite{tegos2020slotted}}&\tabincell{c}{Slotted Aloha}&\tabincell{l}{Throughput}&\tabincell{l}{Apply SIC and joint decoding at an HAP \\to minimize collision}\\ 
\cline{2-5}&
\textbf{\cite{8464255}}&\tabincell{c}{Slotted Aloha}&\tabincell{l}{User fairness}&\tabincell{l}{Optimize base station transmission,\\power, time for energy and data,\\channel access probability,\\transmission rate of each user}\\ 
\cline{2-5}&
\textbf{\cite{pejoski2019rf}}&\tabincell{c}{Slotted Aloha}&\tabincell{c}{Proportional fairness}&\tabincell{l}{Optimize energy harvesting duration,\\channel access probability, base station\\receive power}\\ 
\cline{2-5}&
\textbf{\cite{pejoski2020slotted}}&\tabincell{c}{Slotted Aloha}&\tabincell{c}{Proportional fairness}&\tabincell{l}{Optimize transmit power of users}\\
\cline{2-5}&
\textbf{\cite{lin2020framed}}&\tabincell{c}{Framed Slotted Aloha}&\tabincell{c}{Charging wastage}&\tabincell{c}{Optimize charging order of devices}\\ 
\hline

\tabincell{c}{Power beacon}&\textbf{\cite{iqbal2019learning}}&\tabincell{c}{Framed Slotted Aloha}&\tabincell{l}{Idle or collision slots}&\tabincell{l}{Learn the optimal frame size}\\ [0.5ex]
\hline
                                                                                     
\end{tabular}
}
\end{table*}

%
\subsection{CSMA}\label{sec:CSMA}
Prior works consider using CSMA with Collision Avoidance (CA); aka CSMA/CA, for sending charging requests.  
In particular, the seminal work in~\cite{naderi2014frmac} proposes RF-MAC, where a device with a low energy sends a Request for Energy (RFE) packet to one or more energy transmitters upon sensing that the channel is idle.  These energy transmitters then reply with a pulse, which the device then uses to determine its energy harvesting rate.  The device then determines the best group of energy transmitters and their operating frequency such that the signal from these transmitters adds constructively.  In addition, RF-MAC provides a higher priority to RFE packet and also the device with the highest residual energy when contending for a channel.  Lastly, devices freeze their backoff counter during charging periods.   

Other similar works include \cite{7514670, bi2017distributed, 8390912, 9333647}.
Reference~\cite{7514670} is similar to RF-MAC but the request-charging process is carried out in a centralized manner.
%
Similarly, the devices in \cite{bi2017distributed} send a request, termed Energy Request Buzz (ERB), whenever their energy level is low.  Otherwise, they contend for the channel using $p$-persistent CSMA, where $p$ is set according to the number of devices.
The devices in~\cite{8390912} upon winning the channel, ascertain whether they have sufficient energy to transmit a packet.  If not, they send an energy request packet to a sink, which then supplies the required amount of energy to requesting devices.
Lastly, the protocol in \cite{9333647} introduces a different RTS and CTS message.  Specifically, a device can request no charging before data transmission or request a HAP sends a charge first for some duration.  The energy request probability and charging duration are set as per the energy level of a device.

\begin{figure}[ht]
    \centering
    \includegraphics[width=0.5\textwidth]{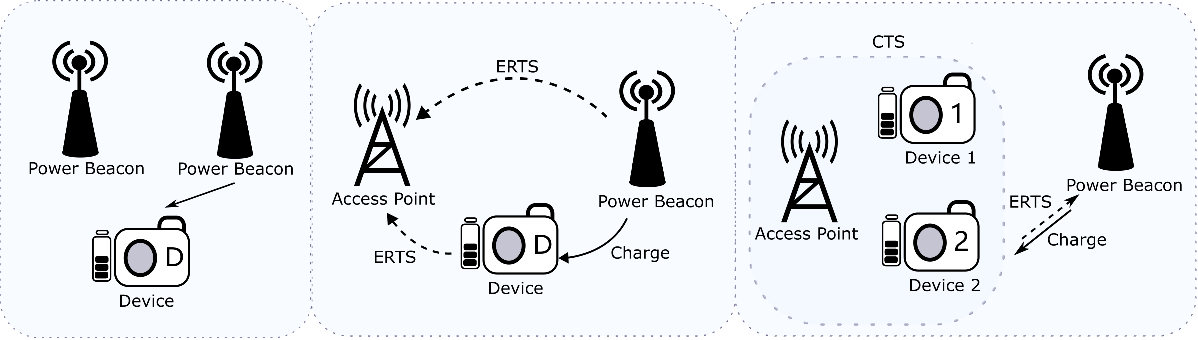}
    \caption{Examples of CSMA works, where power beacons use CSMA/CA to access the channel for charging purposes (left figure).  Alternatively, a device could use an ERTS message, which is routed via its associated access point, to request charging from a power beacon (middle figure).  Lastly, a device could request charging from a power beacon directly (right figure).}
    \label{fig:csma-ca}
\end{figure}

A number of protocols consider power beacons, where they address the key issue of co-existence between their charging signals and data transmissions; note, power beacons operate on the same spectrum as an access point or WiFi networks.
To this end, the protocol in \cite{REACH} is designed for such power beacons, where they use conventional CSMA to access a channel but employ a longer backoff.  Upon winning the channel, they charge devices for some time period that is set according to the idle time of the channel; this ensures the charging process does not interrupt normal data communications.
%
%
In~\cite{8233199}, devices inform power beacons via an access point their energy level.  This information is then used by power beacons to determine when they should contend for the channel and charge devices.  To charge devices, a power beacon sends an Energy Request-to-Send (ERTS) message to the access point after a backoff.  If successful, the access point replies with an Energy Clear-to-Send (ECTS) message.  After that, the power beacon proceeds to charge devices.
In reference~\cite{8896889}, devices request a power beacon to charge them upon hearing a CTS message transmitted by a nearby device. To do so, they set a back-off timer, so called {\em energy backoff}, according to their energy level.  They transmit their request to a power beacon once their energy backoff expires.  Upon receiving a request, a power beacon forms a narrow beam to charge the requesting device.
The aforementioned works are illustrated in Figure~\ref{fig:csma-ca}.

Some works consider an IoT network whereby devices are grouped into clusters, each managed by a HAP.  The HAP has to carry out channel access to charge devices.  Similarly, devices have to carry out channel access in order to transmit to their respective HAP.
In~\cite{8913480}, the HAP of each cluster also acts as a relay, where it forwards data from devices to a base station. Devices are grouped based on their distance to a HAP or their geographical location.   HAPs access the channel using CSMA/CA, where they contend for the channel by sending a RTS message to the base station.  The base station then replies with a CTS message if there is no collision.  The successful HAP then sends a beacon frame to charge its associated devices.   These devices then contend for the channel using framed slotted Aloha.
A related work is \cite{9164981}, where an RF-energy harvesting relay is used to maximize the throughput of a source-destination pair.  Once a source/sender wins the channel, it ascertains whether the path via the relay has sufficient energy and in good channel condition to its destination.  If not, it yields the channel to other senders.
The AP in~\cite{GSMAC} manages multiple clusters of devices; each cluster is supported by a power beacon.  The AP allocates a Restricted Access Window (RAW) to each cluster.  Devices then contend for the channel using CSMA/CA in their allocated window of their cluster.   Successful devices are then scheduled to transmit their data or receive a charge from a power beacon.
Two works have considered charging devices in WiFi networks.
The first is called the harvest-then-transmit-based MAC (HE-MAC)~\cite{ha2018hemac}.   Devices wake-up periodically to receive energy and transmit their data to an access point.  Further, devices have an access priority, where a device with a packet has a higher channel access priority.  At each so called energy harvesting period, the access point transmits an RTS message once the channel is idle.  Devices with a packet then backs off according to their priority.  Once a device's backoff timer expires, it transmits a CTS message.  This causes the access point to transmit energy packets to the device.  After that, the device transmits its data to the access point, if any, and enters sleep mode.   
The access points in \cite{talla2015powering} aim to increase channel utilization.  In particular, a higher channel usage corresponds to higher intensity of RF signals that can be harvested for energy.   In this respect, the access points in \cite{talla2015powering} monitor their queue length, and insert dummy packets when their queue is low on packets.  Consequently, access points are always transmitting a packet or emitting an RF signal. 

There is only one work that combines TDMA and CSMA, which aims to support periodic and aperiodic traffic. Specifically in~\cite{Cho2018}, a superframe has three parts: (i) energy transfer and state information collection, where an HAP charges devices and also collects information such as traffic requirement from devices.  This information is then used to allocate a time slot to devices, (ii) contention-free, whereby devices with an allocated time slot transmit their data, and (iii) contention, whereby devices contend for the channel as per the rules of CSMA.  Further, during idle periods, e.g., all devices have a non-zero backoff counter, the HAP charges devices.

%
\subsubsection{Discussion}
Table~\ref{CSMA_TABLE} summarizes CSMA works, where power beacons may rely on CSMA/CA to access the channel.  In this respect, power beacons or/and devices ensure they do not interrupt existing data transmissions; equivalently, ensure co-existence with legacy data transmitters.   Hence, CSMA plays a critical role in works that aim to support RF charging in existing wireless networks, e.g.,~\cite{talla2015powering}.
Apart from that, devices may rely on CSMA to transmit a charging request to a power beacon or an AP.  Alternatively, in works such as~\cite{\cite{8913480, GSMAC}}, a request corresponds to a permission to access the channel to charge devices and receive data.
\begin{table*}
\Large
\newcommand{\tabincell}[2]{\begin{tabular}{@{}#1@{}}#2\end{tabular}}
\caption{Comparison between CSMA works.}\label{CSMA_TABLE}
  \centering
  \resizebox{\textwidth}{!}{
\begin{tabular}{|c|c|c|l|l|}
\hline
\textbf{Work}              & \textbf{\begin{tabular}[c]{@{}c@{}}Multiple energy\\ transmitters?\end{tabular}} & \textbf{\begin{tabular}[c]{@{}c@{}}Energy transmitters\\ contend for channel?\end{tabular}} & \multicolumn{1}{c|}{\textbf{Aim}}                                                                                                                                                               & \multicolumn{1}{c|}{\textbf{Problem}}                                                                                                                                                           \\ \hline
\cite{naderi2014frmac} & Yes                                                                             & No                                                                                                         & \begin{tabular}[c]{@{}l@{}}Minimize disruption to data transmissions, \\ and maximize received energy at devices\end{tabular}                                       & \begin{tabular}[c]{@{}l@{}}Determine energy transmitters and their\\ energy transfer frequency, charging \\ threshold and duration of sensor nodes\end{tabular} \\ \hline
\cite{7514670} & Yes                                                                             & No                                                                                                         & \begin{tabular}[c]{@{}l@{}}Maximize energy transferred to devices with\\ low energy level, while maximize collected\\ data from devices with a high energy level\end{tabular} & \begin{tabular}[c]{@{}l@{}}Determine devices to receive a charge\\ and charging duration\end{tabular}                                                                              \\ \hline
\cite{bi2017distributed} & No                                                                              & No                                                                                                         & Maximize the sum-rate of devices                                                                                                                                               & Determine operation mode of a HAP                                                                                                                                           \\ \hline
\cite{8390912} & No                                                                              & No                                                                                                         & Decrease packet delivery delay                                                                                                                                             & \begin{tabular}[c]{@{}l@{}}Determine a data transmitter in each\\ slot and whether a sink/HAP transmits\\ energy to a selected transmitter\end{tabular}                      \\ \hline
\cite{9164981} & No                                                                              & No                                                                                                         & Maximize average network throughput                                                                                                                                   & \begin{tabular}[c]{@{}l@{}}Determine whether a user gives up its\\ transmission opportunity\end{tabular}                                                                     \\ \hline
\cite{REACH} & Yes                                                                             & Yes                                                                                                        & Minimize interruptions to data transmissions                                                                                                                                 & \begin{tabular}[c]{@{}l@{}}Determine the length of energy\\ transmission phase\end{tabular}                                                                            \\ \hline
\cite{8233199} & Yes                                                                             & Yes                                                                                                        & \begin{tabular}[c]{@{}l@{}}Minimize RF energy transfer interference\\ to a WiFi station\end{tabular}                                                                     & \begin{tabular}[c]{@{}l@{}}Determine whether power beacons\\ participate in channel contention\end{tabular}                                                                    \\ \hline
\cite{8896889}                 & Yes                                                                             & No                                                                                                         & \begin{tabular}[c]{@{}l@{}}Maximize the throughput and minimize \\ energy wastage\end{tabular}                                                                          & \begin{tabular}[c]{@{}l@{}}Determine charging request probability\\ of stations \end{tabular}                                              \\ \hline
\cite{9333647}                 & No                                                                              & No                                                                                                         & Maximize the throughput of devices                                                                                                                                         & \begin{tabular}[c]{@{}l@{}}Determine the type of RTS message \\ sent by devices\end{tabular}                                                                                 \\ \hline

\cite{8913480}                 & Yes                                                                             & Yes                                                                                                        & \begin{tabular}[c]{@{}l@{}}Minimize intra-group and inter-group\\ collisions\end{tabular}                                                                                  & \begin{tabular}[c]{@{}l@{}}Determine active group for energy and \\ data transfer\end{tabular} 
                                                 \\ \hline
\cite{GSMAC}                 & Yes                                                                             & No                                                                                                         & \begin{tabular}[c]{@{}l@{}}Maximize throughput, delay, and \\ energy efficiency\end{tabular}                                       & \begin{tabular}[c]{@{}l@{}}Divide nodes into energy and data\\ group, and schedule data and energy\\ transfer\end{tabular}                                                   \\ \hline

\cite{ha2018hemac}                 & No                                                                              & Yes                                                                                                        & \begin{tabular}[c]{@{}l@{}}Maximize the energy harvesting rate of each\\ node while satisfying QoS constraints\end{tabular}                                                & \begin{tabular}[c]{@{}l@{}}Coordinate energy transfer of the\\ HAP and data transmissions of nodes\end{tabular}                                                             \\ \hline
\cite{Cho2018}                 & No                                                                              & No                                                                                                         & \begin{tabular}[c]{@{}l@{}}Maximize sum-rate, channel utilization\\ and amount of energy received by nodes\end{tabular}                                                & \begin{tabular}[c]{@{}l@{}}Determine the duration of TDMA and\\ CSMA in each time slot\end{tabular}                                                                         \\ \hline
\end{tabular}
}
\end{table*}
%

%
\section{Contention-Free Based Protocols}\label{CFbasedProtocols}
In this section, devices are assigned a time slot or frequency.
A HAP may poll or grant devices a time slot or frequency based on some criterion.  Further, in addition to a poll message, the HAP may deliver data to devices; see Section~\ref{POLLWORKS} for details.  
Another set of works, see Section~\ref{sec:DTDMA}, consider assigning one or more devices to a time slot; i.e., they construct a TDMA dynamically to meet an objective, e.g., sum-rate.   
Lastly, Section~\ref{sec:NOMA2} reviews works that schedule multiple transmissions to a NOMA-capable HAP or receiver.

\subsection{Polling}\label{POLLWORKS}
A gateway or HAP grants $K$ out of a total of $N$ devices a transmission opportunity via a poll message; this is also known as multi-user scheduling~\cite{C2_trade_off_SWIPT}.
In particular, these $K$ devices/users are chosen to maximize some objective, which includes (i) maximizing sum-rate or throughput, (ii) Age of Information (AoI)~\cite{9380899}, and (iii) fairness. 
There are a number of issues to consider when selecting the best said $K$ devices.  First, obtaining information such as channel gain and energy level of devices may be impractical, especially in large scale networks.  
Second, devices require energy to send and receive data.  Hence, ensuring devices have energy before they are polled is a fundamental issue.
Third, a HAP has to decide whether to use a channel for energy or data transfer.  In other words, a device may receive more energy if the channel condition is favourable.  However, it may not be able to transmit if it is receiving energy, which reduces sum-rate.
In this respect, optimizing over multiple time slots is critical, where a HAP can dedicate a time slot for charging or data transmission, and also make use of future channel gain predictions~\cite{10109156}.
Fourth, an HAP may have to satisfy some quality of service (QoS) requirement, i.e., devices may have a minimum data rate requirement.

\subsubsection{Sum-Rate}
In general, prior works aim to optimize the sum-rate (a) at an HAP, (b) devices, or (c) jointly at both the HAP and devices.  In other words, they aim to select device(s) to maximize the sum rate of uplinks, downlink or both uplinks and downlinks.  
These works are discussed next.

\subsubsection*{Uplinks}
A number of works aim to design a polling protocol that takes into account the recharging rate of devices. 
Example works include~\cite{khan2015zoning, khan2017zoning, misic2014polling}, where a gateway sends out poll messages at some frequency.
For example, in~\cite{khan2015zoning}, devices are grouped into zones according to their distance to a gateway.  The frequency in which zones are polled is dependent on the energy level of devices in a zone.  In this respect, the gateway is assumed to have perfect energy and channel information. 
%
%
The protocol in \cite{khan2015zoning} is then extended in~\cite{khan2017zoning} to consider relays, whereby devices in the zone closest to the gateway acts as relays for devices located in zones farther away from the gateway.
Lastly, in~\cite{misic2014polling, 7031892}, devices are polled in a round robin manner for data.  They may notify a gateway that their energy level is below a given threshold.  Given the set of devices requesting energy, the problem is to find the best policy to charge/receive packet to/from each device.  Specifically, the policy determines when the gateway stops polling for data and starts charging devices.  
%

Some works consider a HAP that uses multiple antennas to full-duplex or beamforming capabilities.
An example work is~\cite{zhai2018accumulate}, where a HAP/scheduler, with full-duplex capability, selects a device with the aim to optimize (i) the average throughput, or (ii) fairness. To achieve (i), it selects the device with the maximum energy to transmit, which may lead to unfairness as a device with a high energy harvesting rate will always be selected by the scheduler.  
This leads to aim (ii), where the scheduler selects devices based on a ratio that is calculated using their current and average energy level. 
In~\cite{C2_throughput12}, the HAP also decides the antennas and their beamforming weight.  Further, the HAP considers the minimum average data rate requirement of devices. 
A key trade-off considered in prior works is whether to use a good channel for information or energy delivery~\cite{C2_trade_off1}.
To this end, the work in \cite{C2_trade_off_SWIPT2} considers the said trade-off by selecting multiple users for transmission subject to their transmission meeting an interference alignment constraint.  Users who are not selected harvest energy. 
Lastly, the authors of~\cite{song2021novel} consider the said trade-off by considering slots that are used for charging or data upload.  Specifically, the HAP in~\cite{song2021novel} schedules charging and data collection mode over multiple time frames with non-causal CSI. To do this, the HAP determines the mode of each time slot that yields the highest system throughput. The authors employ a rolling horizon (RollH) approach coupled with channel gain prediction over a given time window.  Using the same channel gain estimates, the HAP solves an integer linear program over the given window to determine the mode of each time slot.
\subsubsection*{Downlinks}\label{POLLDL}
The works in this category aim to maximize the total data sent from a HAP to devices \cite{C2_QOS1+, C2_QOS2+, C2_QOS3+}. 
A key problem is whether devices should receive data or harvest energy. For example, in \cite{C2_QOS1+}, the problem is (i) to select one device to receive data; other devices harvest energy when the HAP transmits to the selected device, and (ii) to decide the transmit power to each device that maximizes the amount of energy harvested by devices and also ensures their average data rate exceeds a threshold. 
In \cite{C2_QOS2+} and \cite{C2_QOS3+}, a HAP has both grid power and renewable energy.  It aims to (i) select a device to receive data; other devices harvest energy whenever the HAP transmits, and 
(ii) decide the amount of energy to draw from the power grid, subject to a threshold.  The goal is to maximize the average throughput and satisfy the energy requirement of devices.
Note that in \cite{C2_QOS2+}, the fusion center has perfect channel state information of devices, while in~\cite{C2_QOS3+}, it has imperfect channel state information.
Reference~\cite{Rubio2019} considers a setting whereby both the HAP and devices have multiple antennas.  The HAP is responsible for delivering data and energy to devices.  The aim is to optimize the sum-rate at devices over multiple time slots.  A key innovation is a strategy to group devices into two types in each time slot: (i) devices requiring charging, and (ii) devices that receive data.  
To create such groups, the HAP uses the battery level of devices, and greedily adds devices into group (i) in order to maximize sum-rate.  Further, devices are added into group (ii) as long as it does not degrade the sum-rate of devices added to group (i).
%

%
A HAP can be designed to maximize some objective over multiple time slots.
For example, the HAP in \cite{7898453} aims to maximize the weighted sum-rate over $M$ time slots by selecting the device with the highest weight and data rate.  Upon being selected, a device has the choice of using their harvested energy (i) fully, or (ii) partially.   
Lastly, the HAP in~\cite{8516308} uses reinforcement learning to determine a policy to minimize packet loss due to queue overflow.  Specifically, the HAP must select an action which corresponds to the device selected for charging, and providing it with an uplink transmission opportunity.  To do this, it uses the battery level and queue length of devices as state and learn the action that yields the minimal or no packet loss.
%

\subsubsection*{Uplinks and Downlinks}\label{POLLUPDOWN}
A key problem is determining how much channel resource is used for uplinks and downlinks.  In particular, there is the following trade-off: if an HAP uploads more data from devices, then it will have to deliver less data to these devices.
%
In this respect, this trade-off is analyzed  by the authors of \cite{7848950} for a single and multi-user case, where they proposed solutions that determine whether an HAP conducts uplink or downlink mode, and the fraction of time used for downlinks and uplinks.
The HAP in \cite{6884177} uses Time Division Duplex (TDD), where it first transmits data and energy to users.  Then in the uplink period, it assigns a channel to a user.  Its aim is to optimize the energy efficiency of both uplink and downlink transmissions, which is defined as the amount of data transmitted for each Joule of energy.

\subsubsection*{Discussion}\label{Dis_sumrate}
Table~\ref{Sumrate_TABLE} compares works that study polling and aim to optimize sum-rate. In summary, the key problem of prior works is to select one or a set of devices/nodes to transmit/receive data in each time slot according to known information, say energy level, channel state information, or both.  In the case of the HAP having perfect information, a common strategy in past works to improve sum-rate is selecting devices that have good channel and high energy level to transmit/receive data. However, a key limitation of the aforementioned strategy is that devices located further away from the HAP transmit less data to the HAP since these devices normally have worse channel state and lower energy level. To date, only reference~\cite{khan2017zoning} considers using nodes that are located closer to the HAP as relays to improve the amount of data transmitted by further away nodes.  
  Further, a key challenge ignored by most of past works is that is impractical to poll all devices/nodes to obtain battery and channel state information in a large-scale network. There is only one work, i.e., \cite{song2021novel}, that considers the HAP has neither perfect channel state information nor energy state of devices. Specifically, in \cite{song2021novel}, the authors consider channel state prediction instead of directly perfect channel state information.

\begin{table*}
\Large
\newcommand{\tabincell}[2]{\begin{tabular}{@{}#1@{}}#2\end{tabular}}
\caption{Comparison between works that study Polling and sum-rate.}\label{Sumrate_TABLE}
  \centering
  \resizebox{\textwidth}{!}{
\begin{tabular}{|c|c|c|c|c|c|l|}
\hline
\multicolumn{1}{|l|}{\textbf{\begin{tabular}[c]{@{}c@{}}Consider downlink\\ or uplink\end{tabular}}} & \multicolumn{1}{c|}{\textbf{Work}} & \textbf{\begin{tabular}[c]{@{}c@{}}Consider mixed\\ energy source\\ at the HAP\end{tabular}} & \textbf{\begin{tabular}[c]{@{}l@{}}Polled\\ devices\end{tabular}} & \textbf{\begin{tabular}[c]{@{}c@{}}Apply round\\ robin\end{tabular}} & \textbf{\begin{tabular}[c]{@{}l@{}}Multi-hop\\ network\end{tabular}} & \multicolumn{1}{c|}{\textbf{Information}} \\ \hline
\multirow{6}{*}{Uplink} & \cite{khan2015zoning} & No & Multiple & Yes & No & \begin{tabular}[c]{@{}l@{}}Energy level and channel\\ state information of nodes\end{tabular} \\ \cline{2-7} 
 & \cite{khan2017zoning} & No & Multiple & No & Yes & \begin{tabular}[c]{@{}l@{}}Energy level and distance\\ of nodes\end{tabular} \\ \cline{2-7} 
 & \cite{misic2014polling}, \cite{7031892} & No & Multiple & Yes & No & Energy level of nodes \\ \cline{2-7} 
 & \cite{zhai2018accumulate} & No & Single & No & No & Energy level of nodes \\ \cline{2-7} 
 & \cite{C2_throughput12} & No & Single & No & No & \begin{tabular}[c]{@{}l@{}}Channel state information\\ of nodes\end{tabular} \\ \cline{2-7} 
 & \cite{song2021novel} & No & Single & No & No & \begin{tabular}[c]{@{}l@{}}Non-causal channel state\\ information of nodes\end{tabular} \\ \hline
\multirow{12}{*}{Downlink} & \cite{C2_trade_off1} & No & Single & No & No & \begin{tabular}[c]{@{}l@{}}Channel state information\\ of nodes\end{tabular} \\ \cline{2-7} 
 & \cite{C2_trade_off_SWIPT2} & No & Multiple & No & No & \begin{tabular}[c]{@{}l@{}}Channel state information\\ of nodes\end{tabular} \\ \cline{2-7} 
 & \cite{C2_QOS1+} & No & Single & No & No & \begin{tabular}[c]{@{}l@{}}Channel state information\\ of nodes\end{tabular} \\ \cline{2-7} 
 & \cite{C2_QOS2+} & Yes & Single & No & No & \begin{tabular}[c]{@{}l@{}}Channel state information\\ of and energy level of nodes\end{tabular} \\ \cline{2-7} 
 & \cite{C2_QOS3+} & Yes & Single & No & No & \begin{tabular}[c]{@{}l@{}}Imperfect channel state\\ information and energy level\\ of nodes\end{tabular} \\ \cline{2-7} 
 & \cite{Rubio2019} & No & Multiple & No & No & \begin{tabular}[c]{@{}l@{}}Battery level and channel\\ state information of nodes\end{tabular} \\ \cline{2-7} 
 & \cite{7898453} & No & \begin{tabular}[c]{@{}c@{}}Single and\\multiple\end{tabular} & No & No & \begin{tabular}[c]{@{}l@{}}Channel state information\\ of nodes\end{tabular} \\ \cline{2-7} 
 & \cite{8516308} & No & Single & No & No & \begin{tabular}[c]{@{}l@{}}Battery state and queue length\\ of nodes\end{tabular} \\ \hline
\multicolumn{1}{|l|}{\multirow{2}{*}{Downlink and uplink}} & \cite{7848950} & No & Multiple & No & No & \begin{tabular}[c]{@{}l@{}}Channel state information of\\ uplink and downlink nodes\end{tabular} \\ \cline{2-7} 
\multicolumn{1}{|l|}{} & \cite{6884177} & No & Multiple & No & No & \begin{tabular}[c]{@{}l@{}}Channel state of sub-channels\\ and energy level of nodes\end{tabular} \\ \hline
\end{tabular}

}
\end{table*}
%
%

\subsubsection{Age of Information}
\begin{figure}[ht]
    \centering
    \includegraphics[width=0.5\textwidth]{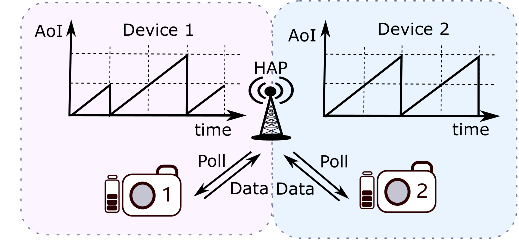}
    \caption{The HAP has to ensure information from sources, namely Device-1 and Device-2, is fresh.}
    \label{fig:poll-aoi}
\end{figure}

Another metric of interest is AoI, which is used to measure the freshness of samples; see~\cite{NET-060} for a survey.  Briefly, assume a destination holds a sample of a time-varying process, which is monitored by a device.  Then AoI measures the elapsed time since the said sample was generated by the device.
In this respect, the goal of prior works is to charge devices, and select devices to transmit a sample in order to minimize AoI.  Figure~\ref{fig:poll-aoi} shows an HAP that collects data from two devices, and the respective AoI for these devices.  In particular, the AoI of a device reduces to zero after the HAP receives an update.  Otherwise, it increases linearly.  In this respect, the HAP's goal therefore could be to minimize the average or peak AoI.

The charging schedule of an HAP has an impact on AoI~\cite{9912162}.
To date, prior works assume the availability of information such as channel gain and energy state/level of devices.
For example, in order to optimize AoI, in \cite{C2_AoI3, C2_AoI7}, and \cite{C2_AoI8}, the scheduler/HAP uses the energy state, uplink channel state, downlink channel state, and the AoI of all devices. Moreover, the work in~\cite{C2_AoI3}, \cite{C2_AoI7}, and \cite{C2_AoI8} decides (i) whether each time slot is used for wireless energy transfer or wireless information transmission, and (ii) the best set of devices to transmit in the wireless information transmission phase to minimize the long-term average weighted AoI. 
There are several differences between \cite{C2_AoI3, C2_AoI7} and \cite{C2_AoI8}. Firstly, in \cite{C2_AoI3}, Jin et al. consider NOMA, which means multiple devices can transmit simultaneously.  On the other hand, in \cite{C2_AoI7}, \cite{C2_AoI8}, the scheduler only selects one sensor to transmit in the wireless information transmission phase. 
Secondly, the solution presented in~\cite{C2_AoI3}, \cite{C2_AoI7}, \cite{C2_AoI8} is different. 
Specifically, in \cite{C2_AoI3}, Jin et al. applied Lyapunov optimization \cite{Lyapunov_optimization} to dynamically decide (i) and (ii) according to the energy state, uplink channel state, downlink channel state, and the AoI of all devices. 
In contrast, the authors of \cite{C2_AoI8} and \cite{C2_AoI7} respectively applied the policy iteration algorithm (PIA)~\cite{Sutton} and deep reinforcement learning (DRL)~\cite{8103164} as a solution.
Another example is reference~\cite{9523861}, where a HAP uses a distributed Q-learning solution to schedule {\em multiple} devices to transmit their sample {\em without} using uplink channel and battery state.   

%
Similar to \cite{song2021novel, 10109156}, the authors of \cite{C2_AoI15b} consider a time slot that can be used for energy transfer or uplink data transmission.   
In this respect, in~\cite{C2_AoI15b}, each uplink data slot has one device selected to transmit data.  Further, there is an intelligent reflective surface, which helps to improve channel condition to/from devices, and a beamforming HAP.  The HAP uses a hierarchical deep reinforcement learning algorithm, where its first layer selects a user and its second layer optimizes the HAP's beamforming weight.
%

A number of works consider an HAP with a directional antenna that charge devices in a geographical area or sector.  In~\cite{9912162}, the aim is to minimize the weighted peak AoI of devices over $T$ time slots.  The key problem is to determine the HAP's antenna orientation to ensure devices receive a charge.  The problem is challenging as there are $M$ sectors or antenna orientations and $N$ devices to choose from in each slot $t$.  Hence, the search space has size $(M\times N)^T$.  In addition, each device may have a different sampling process.   This problem is then studied by the same authors in \cite{9796808}, where they consider the minimum peak AoI and the case with a single and multiple chargers.  In their later work, namely \cite{10188771}, the authors consider average AoI and also optimize the charging duration used by the HAP.
In a different work, the HAP in~\cite{C2_others_AoI} selects a sector to charge, and schedules devices in a sector to upload their data using TDMA.  The HAP's aim is to minimize the energy outage probability of devices, and ensure the AoI of devices is below a threshold. 

Lastly, the work in \cite{10186345} considers scheduling devices over Nakagami-m channels, whereby there is a power beacon that transmits at a fixed power, and an information receiver that maintains the AoI of devices.  Note that charging and information transfer use a different channel.  The goal is to minimize the long-term expected weighted AoI of devices.  To this end, it uses the well-known Lyapunov optimization framework~\cite{Lyapunov_optimization} to schedule a device to transmit at the start of each time slot according to its AoI and harvested energy.  

\subsubsection*{Discussion}
Table \ref{AoI_TABLE} compares prior works that study polling and AoI, where a set of past works consider beamforming \cite{9912162}, \cite{C2_AoI15b}, \cite{9796808}, \cite{C2_others_AoI}. Among these works, references \cite{9912162, C2_AoI15b} and \cite{9796808} jointly optimize data transmission and beamforming, while reference\cite{C2_others_AoI} only focuses on scheduling. Further, most past works designed a centralized algorithm.  By contrast, the nodes in reference \cite{9912162, 9523861}, and \cite{9796808} are able to decide whether to transmit data or a transmission request to the HAP according to their energy level and AoI on their own.   

\begin{table*}
\newcommand{\tabincell}[2]{\begin{tabular}{@{}#1@{}}#2\end{tabular}}
\caption{Comparison between works that study polling and AoI.}
\label{AoI_TABLE}
  \centering
\begin{tabular}{|l|c|c|c|l|l|}
\hline
\textbf{Work} &  \textbf{\begin{tabular}[c]{@{}c@{}} Joint \\ optimization\end{tabular}}& \textbf{\begin{tabular}[c]{@{}l@{}} Polled \\ devices\end{tabular}} & \textbf{Beamforming} & \multicolumn{1}{c|}{\textbf{Information}}                                                                           & \multicolumn{1}{c|}{\textbf{Solution}}                                                                                         \\ \hline                                                                         
\cite{9912162} & Yes                                                          & Multiple                                                               & Yes                    & \begin{tabular}[c]{@{}l@{}} Devices know their energy \\ level  and AoI \\ \end{tabular} & \begin{tabular}[c]{@{}l@{}}Decide transmit devices according \\ to their energy level and AoI\end{tabular} \\ \hline

         \cite{C2_AoI3} & No                                                           & Multiple                                                               & No                     & \begin{tabular}[c]{@{}l@{}}Energy state, channel\\ state and AoI of all\\ devices\end{tabular} & Lyapunov optimization                                                                                     \\ \hline
         \cite{C2_AoI7}   & No                          & Single                                                                          & No                   & \begin{tabular}[c]{@{}l@{}}Channel state, battery\\ state and AoI of all\\ sensors\end{tabular} & \begin{tabular}[c]{@{}l@{}}Deep reinforcement learning\\ algorithm\end{tabular}                           \\ \hline
         \cite{C2_AoI8}   & No                          & Single                                                                          & No                   & \begin{tabular}[c]{@{}l@{}}Channel state, battery\\ state and AoI of all\\ sensors\end{tabular} & Policy iteration algorithm                                                                                 \\ \hline
         & & & & & \\
         \cite{9523861}   & No                          & Multiple                                                                        & No                   & \begin{tabular}[c]{@{}l@{}} Devices know their energy \\ level  and AoI \\ \end{tabular}                                                                                & Distributed Q-learning algorithm                                                                                      \\ 
         & & & & & \\ \hline
\cite{C2_AoI15b}   & Yes                         & Single                                                                          & Yes                  & \begin{tabular}[c]{@{}l@{}}Channel state, energy\\ state, energy demand\\ and AoI of devices \end{tabular} & \begin{tabular}[c]{@{}l@{}}Hierarchical deep reinforcement\\ learning algorithm\end{tabular}              \\ \hline

\cite{9796808}   & Yes                         & Multiple                                                                        & Yes                  & \begin{tabular}[c]{@{}l@{}} Devices know their energy \\ level  and AoI \\ \end{tabular}                                    & \begin{tabular}[c]{@{}l@{}}Each device decides transmission\\ strategy according to its AoI and\\ charging latency\end{tabular} \\ \hline

\cite{C2_others_AoI}   & No                          & Multiple                                                                        & Yes                  & \begin{tabular}[c]{@{}l@{}}AoI, energy level and\\ location of devices\end{tabular}                           & Linear program                                                                                                                  \\ \hline

\cite{10186345}   & No                          & Single                                                                          & No                   & \begin{tabular}[c]{@{}l@{}}AoI and harvested energy\\ of devices\end{tabular}                                 & \begin{tabular}[c]{@{}l@{}}Lyapunov optimization-based online\\ algorithm\end{tabular}                                          \\ \hline
\end{tabular}
\end{table*}

\subsubsection{Fairness}
Ideally, an HAP or fusion center should always poll devices with the best channel or/and energy.  However, this may cause some devices to be neglected by the HAP, namely those that are far from an HAP~\cite{ju2013throughput} or to transmit when they have poor channel conditions.
Hence, balancing throughput and fairness is important.  
In this respect, prior works have proposed various device/user selection strategies to balance throughput and fairness, which includes proportional and equal-throughput fairness.
The first strategy, see~\cite{C2_trade_off_SWIPT}, aims to maximize the average sum-rate subject to the average throughput of users exceeding a given threshold.  The second strategy ensures proportional fairness, where all users have an equal probability to be selected by an HAP.   The third strategy ensures users have equal throughput.
The strategies in~\cite{C2_trade_off1} include (i) selecting the $j$-th user based on their {\em normalized} SNR, which is defined as their instantaneous SNR divided by their average SNR value, or (ii) given an {\em order} set, identifying the users whose normalized SNR belongs to the set.  The scheduler then selects the user with the lowest average throughput.
The work in~\cite{C2_Spectral_efficiency} characterizes the uplink capacity of randomly distributed users.   It studies three strategies to select a user for uplink transmission: (i) selects the device with the best uplink channel condition, (ii) round-robin or conventional scheme that does not consider the energy of users, or (iii) first group users with the minimum required energy to transmit into a set.  Out of these users, a HAP has two strategies: select a user with the best channel or arbitrarily.
Co-channel interference has an impact on the SNR or data rate and energy harvesting rate of users.  Specifically, a user may benefit, in terms of the amount of harvested energy, when there is strong co-channel interference from neighboring base stations.  Conversely, a user should be selected when there is little co-channel interference.  To this end, the work in \cite{C2_trade_off_SWIPT4} outlined an $\alpha$-scheduler, whereby $\alpha\in [0,1]$, trades off the data and harvesting rate of a scheduled user.  Note that the users in \cite{C2_trade_off_SWIPT4} have a power-splitting receiver.
A limitation of the $\alpha$-scheduler is that it does not consider fairness between users.  This is addressed in \cite{C2_trade_off_SWIPT3}, where the metric used to select a user considers the total energy of all users.
%

There is only one work that considers joint uplink and downlink.  
Specifically, in \cite{8294215}, the authors aim to ensure $\alpha$-fairness for both uplink and downlink rate in a WPCN.   The problem is to determine the transmit power of a HAP for data and charging in the downlink part, and schedule the transmission of users/devices in the uplink part.  Further, users employ power splitting to extract data and energy from HAP transmissions.  Users are allocated a pre-defined time slot in the uplink part of a frame.  

Lastly, energy efficiency is a key concern, where the goal is to maximize the amount of data transmitted per unit of energy.  
To date, there are two example works that focus on energy efficiency when selecting a user in each time slot to receive data, whilst other users receive energy.
%
First, the AP in \cite{C2_EE} ensures devices either receive the minimum amount of energy or data rate in each time slot.  
It assigns a sub-carrier to each device for downlink transmission subject to a device's minimum data rate requirement.  As each sub-carrier has a different channel gain to each device, the goal is to assign a sub-carrier in a manner that maximizes energy efficiency.
Similarly, the work in \cite{7572874} also assigns a sub-carrier to devices and optimize transmit power control over sub-carriers. However, the AP in \cite{7572874} optimizes a so called proportional fair energy efficiency metric, which equates to a sum of the logarithm of the data rate received by users.  A key challenge is that the AP has imperfect CSI or experience channel estimation errors.  Note that the work in \cite{C2_EE} assumes the AP has perfect CSI.
\begin{table*}
\newcommand{\tabincell}[2]{\begin{tabular}{@{}#1@{}}#2\end{tabular}}
\caption{Comparison between works that study polling plus fairness or energy outage.}
\label{Fairness_TABLE}
  \centering
  
\begin{tabular}{|l|l|l|l|l|l|}
\hline
\textbf{Works}          & \textbf{SWIPT} & \textbf{TDMA} & \textbf{\begin{tabular}[c]{@{}c@{}} Polled\\ Devices\end{tabular}} & \multicolumn{1}{c|}{\textbf{Information}}                                                                     & \multicolumn{1}{c|}{ \textbf{Solution}}                                                                                                                      \\ \hline
\cite{C2_trade_off_SWIPT}              & Yes            & No            & Single                                                                          & Channel state of each device                                                             & Online algorithm                                                                                                                       \\ \hline
\cite{C2_trade_off1}             & Yes            & No            & Single                                                                          & Channel state of each device                                                             & Online algorithm                                                                                                                       \\ \hline
\cite{C2_Spectral_efficiency}             & No             & No            & Multiple                                                                        & \begin{tabular}[c]{@{}l@{}}Channel state and energy\\ level of each device\end{tabular}  & \begin{tabular}[c]{@{}l@{}}Select devices according\\ to their energy level and\\ greedy strategy\end{tabular}                         \\ \hline
\cite{C2_trade_off_SWIPT4}, \cite{C2_trade_off_SWIPT3} & Yes            & No            & Single                                                                          & \begin{tabular}[c]{@{}l@{}}Channel state and harvested \\ energy of each device\end{tabular}     & \begin{tabular}[c]{@{}l@{}}Select the best device\\ according to achievable\\ data rate and harvested\\ energy at devices\end{tabular} \\ \hline
\cite{8294215}              & Yes             & Yes           & Multiple                                                                        & Channel state of each device                                                             & Iterative method                                                                                                                       \\ \hline
\cite{C2_EE}             & Yes            & No            & Single                                                                          & Channel state of each device                                                             & \begin{tabular}[c]{@{}l@{}}Select devices according to\\ channel state information\end{tabular}                                         \\ \hline
\cite{7572874}             & Yes            & No            & Single                                                                          & \begin{tabular}[c]{@{}l@{}}Imperfect channel state\\ information of devices\end{tabular} & Iterative method                                                                                                                       \\ \hline
\cite{C2_outage1}, \cite{C2_Generalized} & No             & No            & Single                                                                          & \begin{tabular}[c]{@{}l@{}}Channel state and energy\\ level of each device\end{tabular}  & \begin{tabular}[c]{@{}l@{}}Select devices according to\\ known information\end{tabular}                                                 \\ \hline
\end{tabular}

\end{table*}

\subsubsection{Energy Outage}
A key concern is energy outage when a device is given the opportunity to transmit by a HAP.
To date, not many works have considered outage probability and channel access.
In~\cite{C2_outage1} and \cite{C2_Generalized}, there is an energy transmitter that charges $N$ devices, which are then scheduled to transmit to an information receiver.   The goal is to determine a device selection strategy to minimize outage probability.
To this end, the strategies studied in~\cite{C2_outage1, C2_Generalized} include selecting (i) a device randomly, (ii) the device with the best end-to-end signal to noise ratio (SNR); i.e., a device with the best received signal strength from the energy transmitter, and from the device to the information receiver, (iii) the $k$-th device with the highest amount of harvested energy, (iv) the $k$-th device with the best channel to the information receiver, and (v) the $k$-th device with the worst link, which includes both downlink from the energy transmitter and uplink to the information receiver.
These strategies require information in terms of channel gain and/or energy level of devices.
%

\subsubsection*{Discussion}
Table~\ref{Fairness_TABLE} summarizes works that study polling and consider fairness or energy outage. As shown in Table~\ref{Fairness_TABLE}, most of the prior works consider SWIPT. One key problem is to determine the best device to receive data and leave the rest of the devices to harvest energy in each slot so as to achieve fairness \cite{C2_trade_off_SWIPT}, \cite{C2_trade_off1},\cite{C2_trade_off_SWIPT4}, \cite{C2_trade_off_SWIPT3}, \cite{8294215},\cite{C2_EE},\cite{7572874}. Only one work joint considers SWIPT and TDMA \cite{8294215}. Specifically, in\cite{8294215}, instead of selecting the best device in each frame, the key problem is to assign a duration for each device to receive data/energy and transmit data. We notice that one common characteristic of prior works is that all of them require channel state information and/or energy level of devices to optimize device selection/time allocation, which is impractical in a large-scale network.

\subsection{Dynamic TDMA}\label{sec:DTDMA}
There are limited works that consider assigning one or more devices to a transmission slot dynamically.
Note that most WPCN works pre-assign devices to a {\em fixed} transmission slot.  Once devices are assigned, the main problem is then to optimize the (i) size of time slots for charging or/and data transmissions, and/or (ii) transmit power of devices; see~\cite{kang2015full, terauchi2021harvest} for some examples.  This means they do not have a channel access problem; these works are omitted from our discussion.
By contrast, we consider works that aim to assign a device to a time slot, aka user scheduling.  
The time slot assigned to a device has an impact on their available energy.  That is, devices assigned to a later time slot have more opportunities to harvest energy; equivalently, a device with low energy should transmit last; this allows the device to have a higher data rate as it will have more time to accumulate energy~\cite{kang2014optimal}.
To this end, a number of works have studied different transmission orders.
For example, in~\cite{YoonTDMA}, the authors study how slot allocation affects sum-rate and fairness.  They proposed an ordering that uses the SNR of users, where they showed that a descending SNR order leads to better fairness whilst an ascending SNR order results in a higher sum-rate.  This ordering is also considered in~\cite{kang2014optimal} for a full-duplex HAP.
A challenging case is considered in \cite{9162112}, where an HAP creates an uplink transmission schedule without using the energy level of devices.  Specifically, after each charging time slot, the HAP creates a TDMA schedule for one or more devices.  As the HAP is unaware of the energy level of devices, some time slots may be idle.   Note that the number of selected devices affect the throughput of devices.  This is because a device that is not selected in a frame have to be scheduled in a future frame.  Hence, the key problem in \cite{9162112} is to decide on the sets of devices after each charging time slot such that devices transmit frequently, and the resulting schedule has minimal or zero idle time slots.

The HAP in~\cite{10077429} also considers data collection time.  To this end, they address two problems: (i) maximize throughput subject to a time budget, and (ii) minimize data upload time subject to each device uploading a minimum amount of data.
For problem (i), the optimal schedule places devices with a high channel gain or low energy consumption to transmit as late as possible so that they can accumulate more energy, and thus they will have a high data rate.  
For (ii), the device with a larger minimum amount of data requirement is scheduled later.
Lastly, in \cite{iqbal2022minimum}, users have a minimum data rate requirement over some planning horizon.  To this end, the authors consider the problem of assigning users to a time slot, and minimizing the data upload time duration of users.  Further, users have to select a data rate and set their transmit power.

To date, only two works have built upon a standardized protocol. 
First, the work in~\cite{s22124520} considers a Zigbee or IEEE 802.15.4e~\cite{2006wireless} cluster-tree network, where each cluster has a HAP and RF-charging devices.  They extend the superframe structure of IEEE 802.15.4e to allow the respective HAP of clusters to charge devices, and support different traffic types. 
Second, the protocol in~\cite{electronics10010048} allows devices to request for transmission opportunities in a slotframe of IEEE 802.15.4 to facilitate energy delivery and timely delivery of data.  Specifically, the protocol determines the number of cells\footnote{Each cell has a time and channel offset in each slotframe.} in slotframes required for energy delivery and data transmissions.  It then uses the 6P protocol \cite{rfc8480} to reserve these cells.  The protocol in~\cite{electronics10010048}, however, requires devices to switch channel frequently.  This problem is addressed in \cite{9895109} whereby devices are allocated cells on the same channel.
%

In summary, prior works have studied how transmission order impacts throughput and fairness, e.g., ~\cite{kang2014optimal, YoonTDMA}.   If the goal is throughput maximization, then devices with a good channel must be scheduled later in time to allow for a higher data rate or energy level.  Otherwise, if fairness is an issue, then devices with a good channel are to be scheduled first or earlier.  Further, prior works have considered a transmission schedule that ensures devices upload a minimum amount of data, i.e., \cite{iqbal2022minimum}.  
A key limitation of these works, however, is that they assume perfect knowledge of energy level at devices.  To date, only the work in \cite{9162112} has considered creating a dynamic TDMA schedule without using the said information.  It further considers battery leakage.
Apart from that, in~\cite{electronics10010048}, periodic and aperiodic traffic types are considered.  Further, only references~\cite{electronics10010048, s22124520} have considered assigning time slots for energy delivery.

\subsection{NOMA}\label{sec:NOMA2}
\begin{figure}[ht]
    \centering
    \includegraphics[width=0.5\textwidth]{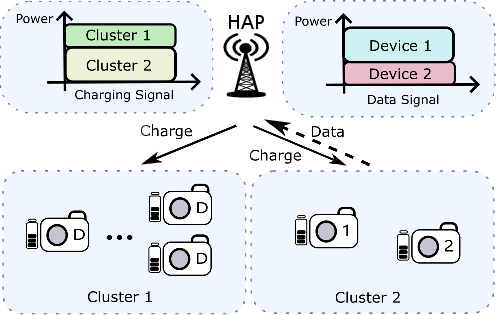}
    \caption{A NOMA-based energy harvesting network.}
    \label{fig:cluster-noma}
\end{figure}

Prior NOMA works assume devices/users are allocated to a time slot~\cite{islam2017power, wang2020throughput} or assume a {\em fixed} decoding order as per their channel gain~\cite{dia2016wireless}.  Hence, these works are omitted from our discussion.
Instead, our focus is on works that involve device/user selection; equivalently, we outline works whereby the set of transmitting devices/users depend on network/node/channel condition.
Note that the decoding order at a receiver has a significant impact on performance~\cite{islam2017power}.  However, determining the best order requires an exhaustive search, meaning it becomes computationally intractable for large number of devices/users~\cite{6512494}.
Another key issue is signaling overheads relating to obtaining channel gain information from devices~\cite{9739685}; note, most works assume this information is readily available and thus decoding order is known.  In practice, obtaining the said information becomes expensive in large-scale IoT networks.  Further, channel estimation uses the precious harvested energy of devices that could have been used to improve data rate.
In many works, grouping devices is a fundamental problem, where devices in the same group can transmit simultaneously to a SIC-capable HAP.  
Figure~\ref{fig:cluster-noma} shows a HAP serving two groups/clusters of devices/users.  The HAP may charge each group in turn or all devices simultaneously.  Further, devices transmit to the HAP using NOMA, and vice-versa.

To form such a group, the work in~\cite{CNOMA2020a} constructs groups by ensuring devices in the same group  have the widest channel gain difference; this helps facilitates SIC decoding~\cite{7273963, 9013805}.  Further, in \cite{7273963}, each cluster is assigned a distinct frequency, whereas in \cite{9013805}, it is given a time slot.  
%
A similar strategy for group construction is outlined in \cite{9599545, 9563077}, where the authors assume two sets of users, say $S_A$ and $S_B$.  Users in set $S_A$ are located closer to a HAP.  Then a pair of transmitting users is selected, where they propose to pair users in set $S_A$ and $S_B$ based on their channel gain difference to help facilitate SIC decoding.  Further, they select a user in set $S_A$ that maximizes energy efficiency.  
%
%

In \cite{8485639}, the authors consider a millimeter massive MIMO-NOMA network where users employ power splitting.  In this setting, the authors propose to group users according to hybrid precoding and beamforming.  Specifically, for each beam, the proposed algorithm starts with the device that has the strongest channel gain; this device becomes the cluster head.  Other users with channel correlation that is less than a given threshold are grouped together with the cluster head.
%
A device's data rate is affected by the number of concurrently transmitting devices in a group.  To this end, the work in \cite{NOMASDMA1} outlines an algorithm that employs Spatial Division Multiple Access (SDMA) to service different sectors.  Users in each group are then paired together to facilitate SIC decoding.

The work in \cite{8761702, 9846219, 9096318} considers the doubly near-far problem~\cite{ju2013throughput}.  To overcome data rate unfairness, they consider a system setup with a power beacon that charges devices, and an access point that receives data from devices.  Further, the aforementioned works use a time frame with the following phases: (i) charging, (ii) charging and data upload, (iii) data upload only.  To determine data transmissions, they place devices with good channel condition into so called {\em interference} group.  Other devices are placed in a {\em non-interference} group.  Each device in the {\em interference} group is assigned a dedicated time slot in phase (ii) of the frame.  Note that the power beacon transmits in both phase (i) and (ii).  Further, observe that devices in the {\em non-interference} group receive more charging time.  This helps overcome the doubly near-far problem as these devices have more energy to transmit at a higher rate.  Lastly, each device in the {\em non-interference} group is assigned a time slot in phase (iii).  
The same authors consider max-min fairness in \cite{8761702}.  
After that, in \cite{8761702}, they propose schemes that group devices in phase (iii) so that they can transmit concurrently to an access point.  More specifically, they proposed two schemes that determine whether devices are paired first based on their channel gains before they are grouped together, or devices are grouped together first and then paired together to transmit.

A challenging setting is assigning devices/users to a time slot using only imperfect CSI.  This means they do not assume devices are located at a given position, i.e., far or near a HAP/receiver.
In this respect, there are only two works.
The authors of~\cite{liu2020link, 8972552} investigated a link scheduling problem that aims to maximize the expected throughput at an HAP. The HAP schedules devices into one or more data slots and decodes data using SIC. The problem is thus to find the optimal transmission schedule over time without using any CSI. 
%
Reference \cite{9745145} considers a system with a full-duplex HAP with multiple antennas and random channel gains. It aims to maximize the average sum-rate over multiple time slots.  Further, devices have a minimum uplink data rate requirement and transmit with a fixed power.  The main variables to be optimized at the HAP include the antennas used for energy transfer and data reception, beamforming weight of antennas used for energy delivery, the device scheduled to transmit in each time slot.  Here, the HAP selects the device with the highest data rate. 
%

Lastly, reference~\cite{10109156} considers a HAP that uses a novel time structure; each slot is used for charging or data transmission from devices; see Figure~\ref{fig:timeslot3}.  The goal is to maximize the number of packets received by the HAP over $T$ time slots.  The problem is to determine how many and which of these $T$ time slots are used for charging and data transmissions.  Moreover, in data time slots, a HAP has to select a set of devices to upload their data.  The problem is challenging as there are many time slots configurations, and each data time slot has a different decoding order or combination of devices that can upload to the HAP successfully.


%
Table~\ref{Table NOMA} summarizes NOMA works.  The general idea is to pair users with different channel gains together; the main criterion is to ensure concurrent transmissions to a receiver/HAP have sufficiently different received power.  In this respect, prior works either assume users are already placed into groups, e.g.,~\cite{9013805, CNOMA2020a}, or a pair of users is selected in an on-demand manner, e.g., when a user wants to transmit, another user that allows SIC decoding is selected~\cite{7273963}.
Most works consider uplinks from users.  Only references \cite{8485639} and \cite{9013805} have considered downlinks.  Further, fairness is an important issue as a user/device may be neglected if it is not selected frequently into a group, see \cite{CNOMA2020a, NOMASDMA1}.
Lastly, except for~\cite{liu2020link, 8972552, 9745145}, all works assume perfect channel gains to/from users.  Further, except~\cite{10109156}, all works optimize over a single time slot.
%

%
\begin{table*}
\Large
\newcommand{\tabincell}[2]{\begin{tabular}{@{}#1@{}}#2\end{tabular}}
\caption{Comparison between NOMA-based communication networks. Here, UL and DL denote uplinks and downlinks, respectively.}
\label{Table NOMA}
  \centering
  \resizebox{\textwidth}{!}{
  \begin{tabular}{|c|c|c|c|c|c| }
\hline
\textbf{Works} & \textbf{System}&\textbf{Channel Access (UL/DL)}&\textbf{Key Points}&\textbf{Aim}&\textbf{Problem}\\ 
\hline
\hline
\textbf{\cite{CNOMA2020a}}& {\tabincell{c}{An HAP, clustered\\ users}}&\tabincell{c}{\\ NOMA, OFDMA, (UL) \\}&\tabincell{l}{Each group has a time allocation}&\tabincell{l}{Improve fairness between\\ users, and maximize the\\ uplink sum rate }&\tabincell{l}{Optimize transmit power of \\users, and time allocated to groups}\\ [0.5ex]
\hline
\multirow{2}{*}{\textbf{\cite{7273963}}}&
\multirow{2}{*}{\tabincell{c}{A cellular\\communication system\\with a base station}}
&\tabincell{l}{F-NOMA (UL)} & \tabincell{l}{Order uses according to\\their channel power gain} & \tabincell{l}{Sum-rate}&\multirow{2}{*}{\tabincell{l}{User selection, and\\ transmit power control}}\\
\cline{3-5}
&{}&\tabincell{l}{CR-NOMA (UL) }&\tabincell{c}{Pair the user with a strong channel\\and one with a poor channel condition}&\tabincell{l}{Sum-rate and guarantee\\ data rate of paired user}&\\ [0.5ex]
\hline
\textbf{\cite{9013805}}& {\tabincell{c}{An HAP with\\ clustered users}}&\tabincell{c}{TDMA-NOMA (DL) }&\tabincell{l}{Allocate time slots for \\different user groups}&\tabincell{c}{Minimize transmit power\\subject to minimum data rate\\ and energy harvesting rate}&\tabincell{l}{Determine power allocation and\\power splitting ratio}\\ [0.5ex]
\hline
\textbf{\cite{9599545}}& \tabincell{c}{An HAP, users located\\ in near and far field}&\tabincell{l}{NOMA (UL) }&\tabincell{c}{Pair users from near and far fields\\to achieve high channel gain difference}&\tabincell{c}{Optimize energy efficiency \\and spectral efficiency}&\tabincell{l}{Determine users grouping and\\power allocation}\\ [0.5ex]
\hline
\textbf{\cite{9563077}} & \tabincell{c}{\\An HAP, users located \\in near and far field\\ \\}&\tabincell{l}{NOMA (UL) }&\tabincell{c}{Pair users from near and far fields\\to achieve high channel gain difference}&\tabincell{l}{Maximize sum-rate of users}&\tabincell{c}{Determine time allocation \\for energy transfer}\\ [0.5ex]
\hline
\textbf{\cite{8485639}} & \tabincell{c}{A single-cell downlink\\mmWave massive\\MIMO-NOMA system\\ }&\tabincell{l}{NOMA (DL) }&\tabincell{c}{Group users according to hybrid\\precoding and beamforming }&\tabincell{l}{Maximize sum rate}&\tabincell{c}{Cluster head selection, transmit\\ power allocation, and power \\splitting factor}\\ [0.5ex]
\hline  %
\textbf{\cite{NOMASDMA1}}& \tabincell{c}{A multi-antenna HAP and\\ paired users}&\tabincell{l}{SDMA and NOMA (UL)}&\tabincell{c}{Users are divided into sectors/areas and\\ served using SDMA.  NOMA is used\\ for users in each sector.}&\tabincell{c}{Maximize sum-rate, and \\ fairness}&\tabincell{l}{Determine the number of sectors}\\ [0.5ex]
\hline
\textbf{\cite{8761702, 9846219, 9096318}}& \tabincell{c}{A power beacon,\\ and an AP}&\tabincell{l}{NOMA (UL)}&\tabincell{c}{Group randomly deployed users\\ based on channel conditions}&\tabincell{l}{Max-min rate}&\tabincell{l}{Grouping of users into time slots}\\ [0.5ex]
\hline
\textbf{\cite{liu2020link, 8972552}}& \tabincell{c}{An HAP, \\ and devices}&\tabincell{l}{NOMA (UL)}&\tabincell{c}{Construct a TDMA link schedule with\\ imperfect channel information}&\tabincell{l}{Sum-rate}&\tabincell{l}{Constructing a TDMA schedule}\\ [0.5ex]
\hline
\textbf{\cite{10109156}}& \tabincell{c}{An HAP and\\ devices}&\tabincell{l}{NOMA (UL)}&\tabincell{c}{Employ a mode-based time structure to\\ take advantage of channel condition\\conducive of charging or data transmissions}&\tabincell{l}{Sum-rate}&\tabincell{l}{Select mode (charge or data), and\\ device selection in data time slots}\\ [0.5ex]
\hline
\end{tabular}
}
\end{table*}

%
\begin{figure*}[tp]
\centering
\subfloat[] 
{\includegraphics[width=.50\linewidth]{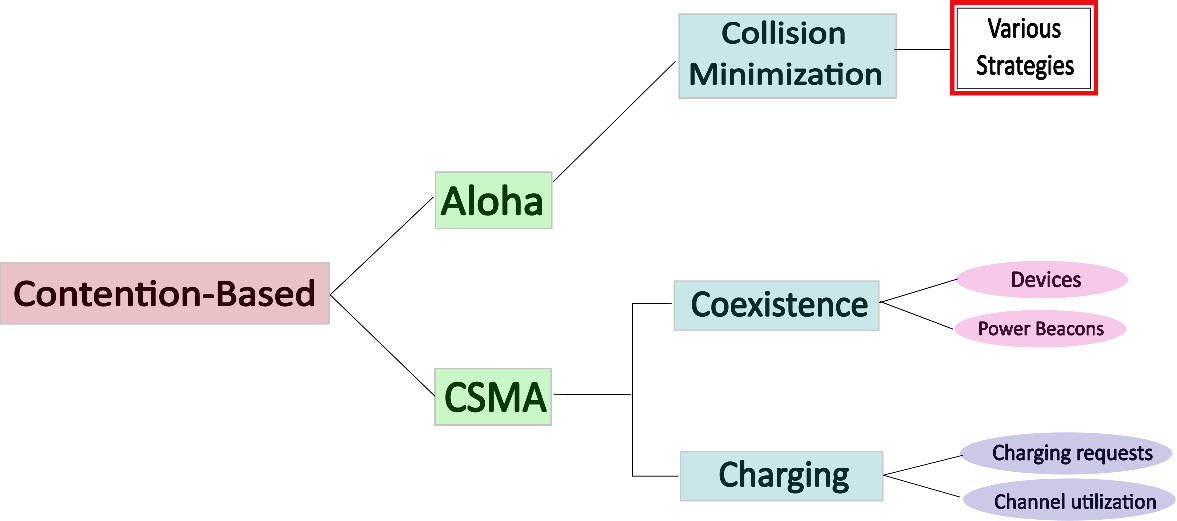}\label{fig:da1}}\hspace*{-0.0em}
\hfill
\subfloat[]
{\includegraphics [width=.450\linewidth]{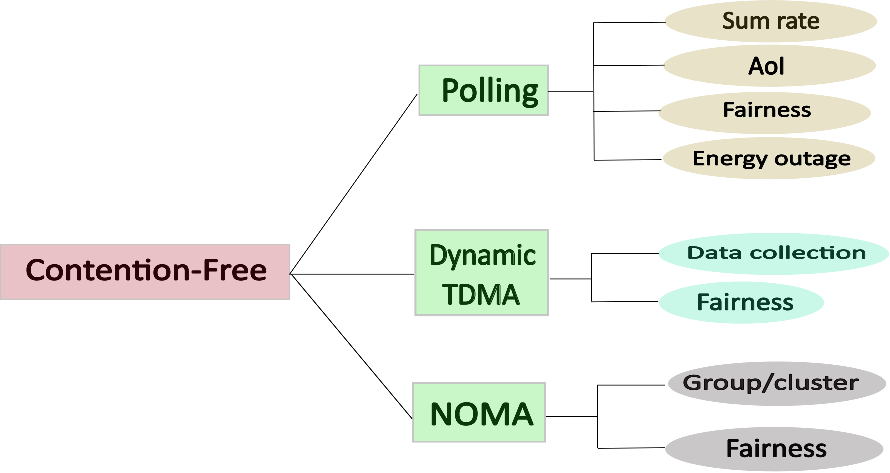}\label{fig:da2}}\hspace*{-0.0em}
\caption{Key research aims of works that target contention-based and contention-free channel access protocols. For 'various strategies', please refer to Figure~\ref{fig:cstrat} and its accompanying discussion. }
\label{fig:ddis}
\end{figure*} 

\section{Discussion}
Figure~\ref{fig:ddis} compares the research aims of works that designed/studied contention-based and contention-free protocols.
In general, both set of works consider fairness in terms of throughput or sum-rate, which stems from the double near-far problem~\cite{ju2013throughput}.   
Another common aspect is some works use SIC to minimize collision or/and improve throughput. In this respect, channel access protocols have to consider grouping of devices or/and optimize the transmit power of devices to facilitate SIC decoding.
A strong assumption in contention-less works is that a HAP has perfect channel gain information or knowledge of energy level at devices.  This information then allows the HAP to group devices or/and set the optimal transmit power to ensure concurrent transmissions have the widest possible received power.   Note that the transmit power of a device is a function of its energy level, which is coupled to the transmit power and duration used by a HAP or power beacon.  In addition, it also depends on a device's past energy usage and channel gains.
Another observation is that very few works, namely~\cite{iqbal2019learning, li2021random, 9632809, 8516308}, have employed reinforcement learning along with dedicated RF sources to optimize channel access protocols.
Lastly, except~\cite{9615376, 9145865, 8049272, 8913480, GSMAC, 8913480, GSMAC, khan2017zoning},  most works focus on single-hop networks.  As for multi-hop networks, they have only considered one or more relays; i.e., prior works have only studied two-hop networks.
%

\begin{figure}[htbp] 
    \centering 
    \includegraphics[width=0.40\textwidth]{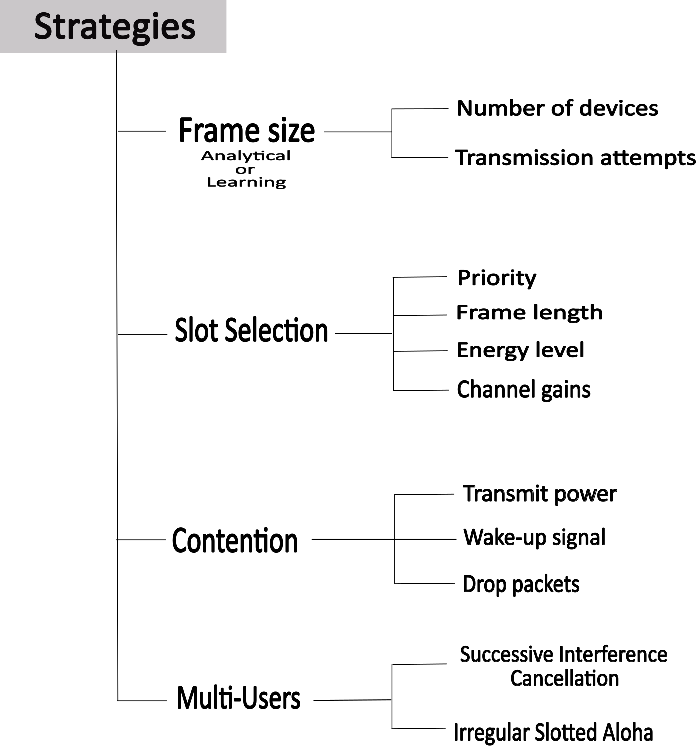} 
    \caption{Strategies used by contention-based protocols to minimize collisions.}
    \label{fig:cstrat} 
\end{figure}

In general, works that employ contention-based protocols aim to minimize collision; see Figure~\ref{fig:da1}.  Unlike conventional wireless networks, the contention level of a channel relates to the number of devices with sufficient energy to transmit a packet.  
%
In this respect, various strategies are used to adapt channel access parameters; see Figure~\ref{fig:cstrat}.
In the {\em first} strategy, a HAP varies its frame size based on the number contending devices, and transmission attempts; both information can then be used by the HAP to derive analytically or via learning, e.g., \cite{iqbal2019learning, li2021random}, the optimal frame size that has the highest transmission success.  

The {\em second} strategy is to set different transmission priorities, where devices far away from a HAP transmit later or they may not transmit if their channel is poor, e.g.,~\cite{choi2019harvestuntil}. The transmission probability of devices can also be set according to their energy level; equivalently their distance to the HAP.  Moreover, devices may only transmit if they have energy to ensure a given received power at the HAP.  Lastly, different frame sizes also induce different collision or success probability.  For example, in \cite{yu2019two}, devices change their transmission probability depending on the frame size advertised by the HAP.

The {\em third} strategy is to reduce the number of contending devices.  A HAP achieves this by changing its transmit power, charging frequency or/and duration, e.g., \cite{silva2021slotted}; all of which affect the number of devices contending for a channel. Another option, see \cite{silva2021slotted}, is to remove packets from devices  
Lastly, a HAP can be equipped with multi-user detection technologies~\cite{multiuser1998verdu}, e.g., SIC, which allows it to decode concurrent transmissions.  This in turn affects the frame size and transmission probability used by a HAP and devices, respectively.  
As for CSMA, prior works aim to schedule charging requests by devices or power beacons.  Specifically, a device may use CSMA/CA to send a charging request to one or more RF energy sources.  Similarly, power beacons may use CSMA/CA to reserve the channel to transmit energy packets.   Another example is~\cite{8913480} where cluster leaders use CSMA/CA to reserve the channel so that cluster members/devices may access the channel.  
A critical issue is ensuring RF charging does not interfere with data transmissions.  Similarly, power beacons use CSMA/CA to co-exist with legacy devices or allowing them to operate in existing WiFi networks.
In this respect, an access point may ascertain the utilization of a channel, see ~\cite{talla2015powering}, and transmit additional packets to increase utilization.  
Alternatively, works such as~\cite{choi2019harvestor, Cho2018} transmit packets opportunistically when a channel is idle.  Both efforts ensure devices have a high RF energy harvesting rate.
%

%
On the other hand, for contention-free protocols, see Figure~\ref{fig:da2}, the general aim is to select or schedule devices into a given set of time slots.  
Specifically, a HAP can poll a device for data or to deliver data.  In this respect, prior works have aimed to maximize the sum-rate of uplinks, downlinks or both.   A main issue is to ascertain whether a device has sufficient energy or/and good channel condition when it is polled by a HAP.
For works that consider downlinks, a key issue is to determine which devices or areas are to be charged by a HAP or to receive data.  Further, a HAP has to optimize its transmit power or antenna weights to ensure the data rate of the device receiving data is above a given requirement, and at the same time, deliver sufficient energy to other devices.
This issue also arises in works that aim to minimize AoI.  Further, these works have considered an HAP with SIC capability.

In terms of dynamic TDMA, the order of transmissions or schedule has an impact performance, e.g., throughput.  If throughput is of interest, then a HAP should place devices with a good channel condition at the end of a schedule.  Otherwise, to account for the doubly near-far problem or devices far from an energy source, then these devices should be allowed to spend more time harvesting energy. 

Lastly, for NOMA works, the HAP can derive a transmission schedule that consists of one or more transmitting devices.  That is, if the HAP has SIC capability, then a key problem is grouping devices together to take advantage of multi-user detection.  To do this, prior works generally use the channel gain of devices, where they aim to ensure devices in the same group have a significantly different received power.
In all categories of channel access protocols, many works have addressed fairness, whereby they ensure far away devices or those with a poor energy or channel condition transmit at a later time or ensure devices have a minimum data rate.


%
%
\section{Future Works}
We now list some possible future research directions that aim to develop channel access protocols for RF-energy harvesting IoT networks.  They are as follows:
\begin{itemize}
    \item Graph Neural Networks (GNNs) are ideal for modeling communication networks~\cite{9618652}, and conflict graphs or interference~\cite{9417216}.  One possible research direction is to consider the energy availability of devices, where a link between devices exist if they have sufficient energy to communicate.  In this respect, a GNN can be used to predict these graphs or links for different HAP transmit power, charging frequency, duration, or the number of power beacons.  With a predicted graph in hand, we can then compute a transmission schedule. 
    \item To facilitate channel access, digital twins~\cite{9120192} of sensor devices or an RF-energy harvesting IoT network can be created to study different channel access and charging parameters. 
    For example, digital twins of devices can be used to simulate their energy arrivals over time, e.g.,~\cite{10302351}.  This information can then be used to set the size of a frame, transmission probability of devices, transmit power and set forth.    
    \item IRSA thus far has received little attention.  There are works that assume ambient an energy source, see \cite{demirhan2018energy}.  However, apart from \cite{9632809}, there is no other IRSA works that consider a dedicated RF source.  Therefore, there is an opportunity to design protocols whereby RF sources must provide sufficient energy to devices in order to transmit one or more packet replicas depending on the contention level of a channel.

    \item Considering technologies such as IEEE 802.11 (WiFi) or IEEE 802.15.4 (Zigbee) is important as any developed protocols will be able to work in currently deployed wireless networks.  A research direction is therefore to adopt existing standardized channel access protocols to ensure RF-energy harvesting devices are able to co-exist with legacy/conventional non-RF energy harvesting devices.  For some works in this direction, the reader is referred to works such as \cite{s22124520, electronics10010048, ha2018hemac,  talla2015powering}.
    \item Recently, Intelligent Reflecting Surfaces (IRS)~\cite{9326394} has received considerable attention by the physical layer community.  However, the networking community has only begun to embrace its full potential; for more information, the reader is referred to \cite{9385372} and \cite{9475159}, their references and citations.
    In this respect, a research direction is to design IRS-aided channel access protocols to facilitate efficient charging, concurrent data transmissions, and/of co-existence of charging and data transmissions.   
    %
    \item There is little research on joint uplinks and downlinks optimization.  Moreover, works such as~\cite{8294215} have only considered an HAP that employs a polling protocol.  However, no others works that design non-polling based channel access protocols exist, where these protocols must optimize the RF energy of devices for data reception, and perhaps processing, and data uploads. 
    \item To date, no works have considered devices that use CSMA/CA and a receiver or HAP that has SIC capability.  We note that works such as \cite{7218584} have proposed adopting CSMA to take advantage of SIC.  However, these works do not consider devices that have varying energy levels over time or require charging before transmissions.      
    Further, in this respect, no works have considered Rate Splitting Multiple Access (RSMA)~\cite{9831440}.
    \item In terms of reinforcement learning, a possible research direction is to assume devices and power beacons contain an agent that collaboratively learn channel access parameters such as frame length and transmission probability, in order to optimize some objective.    Note that works such as \cite{9435617} have considered multi-agent reinforcement learning (MARL) in an IoT system that uses random access.  However, no works have considered applying MARL to improve the channel access in an energy harvesting IoT system.
    A key concern is that devices may not have sufficient energy to train an agent.  In this respect, a possible future work is to consider platform such as IoT sensor gym~\cite{SensorGym}.
    Lastly, another possible use of reinforcement learning is to determine the type of channel access protocol to apply in a given scenario.  One possible direction is to extend the work in~\cite{8665952} to consider dedicated RF-energy sources.  
    \item A key challenge when using contention-free protocols in large-scale IoT networks is gathering channel gain information.  However,  this process requires devices to use their energy to process and reply to pilot symbols emitted by a HAP or power beacon~\cite{9739685}.  In this respect, there is a dearth of works that assume imperfect channel gain information; see \cite{C2_QOS3+, liu2020link, 8972552, 9162112}; this direction is important for practical purposes.   Hence, there are ample opportunities to revisit existing works but consider the imperfect channel gain or energy level information case.
    
    \item Similarly, only a few works have considered optimization over multiple time slots, see \cite{yu2021data, song2021novel, 10109156} and Figure~\ref{fig:timeslot3}, where a node takes advantage of future channel gains for charging or/and data transmissions.   Possible works include designing channel access protocols that leverage reinforcement learning, digital twins or/and model predictive control as a solution framework.
    \item Another possible direction is to leverage the Predict-n-Optimize framework~\cite{SPO}.  Its main aim is to use machine learning techniques coupled with the optimization model of a system to minimize {\em decision} errors.  It is worth emphasizing that Predict-n-Optimize is distinct from works that use machine learning to predict the coefficients/data of an optimization model.  As an example, works such as~\cite{9372291} use a neural network to predict energy arrivals and channel gains, where the estimates of these quantities are then used to optimize the transmit power of devices.   Hence, for works such as~\cite{9372291}, prediction errors are of key concern.
    By contrast, the Predict-n-Optimize framework~\cite{SPO} focuses on {\em decision} errors, where some data, e.g., solar irradiance of an area or channel utilization of a network, is used to compute a decision, which could correspond to a HAP polling a device or frame size; in this respect, an incorrect decision may cause the wrong device to receive a poll message or a frame size that results in excessive collision or idle slots.
%
\end{itemize}

%
\section{Conclusion}
This paper has presented the first comprehensive survey on channel access methods/solutions that assume one or more {\em dedicated} RF-energy sources.  It has discussed contention-based protocols such as Aloha, and contention-free protocols such as polling, where these protocols are used to schedule transmissions of data or energy delivery messages.
It has analyzed both categories of protocols and distill their key ideas, aims, and performance metric(s).
Lastly, this survey has outlined a number of interesting research directions, which include the use of machine learning techniques or/and digital twins for decision making, and designing protocols that leverage intelligent surfaces.
%

\bibliographystyle{ieeetr}
\bibliography{Refs}
\end{document}